\documentclass[AMA,STIX1COL]{WileyNJD-v2}

\articletype{Research Article}%
\usepackage{graphicx, algorithmicx, setspace, amsmath, amssymb, subfigure, url, multirow, booktabs, verbatim, algpseudocode, bm,threeparttable,algorithm, color, amsfonts}
\usepackage{amsthm, rotating,mathtools}

\received{}
\revised{}
\accepted{}

\begin{document}

\title{Biomarker-guided heterogeneity analysis of genetic regulations via multivariate sparse fusion}

\author[1]{Sanguo Zhang}

\author[1]{Xiaonan Hu}

\author[2]{Ziye Luo}

\author[3]{Yu Jiang}

\author[2]{Yifan Sun*}

\author[2,4]{Shuangge Ma*}

\authormark{Sanguo Zhang \textsc{et al}}

\address[1]{\orgdiv{School of Mathematical Sciences, and Key Laboratory of Big Data Mining and Knowledge Management}, \orgname{Chinese Academy of Science}, \orgaddress{\state{Beijing}, \country{China}}}

\address[2]{\orgdiv{Center of Applied Statistics, School of Statistics}, \orgname{Renmin University of China}, \orgaddress{\state{Beijing}, \country{China}}}

\address[3]{\orgdiv{School of Public Health}, \orgname{University of Memphis}, \orgaddress{\state{Tennessee}, \country{USA}}}

\address[4]{\orgdiv{Department of Biostatistics}, \orgname{Yale University}, \orgaddress{\state{Connecticut}, \country{USA}}}

\corres{*Yifan Sun, No. 59 Zhongguancun Street, Haidian District, Beijing, 100872 \\\email{sunyifan@ruc.edu.cn} \\
*Shuangge Ma, 60 College ST, New Haven, CT, 06520, U.S.A. \\
\email{shuangge.ma@yale.edu}
}

\abstract[Summary]{Heterogeneity is a hallmark of many complex diseases. There are multiple ways of defining heterogeneity, among which the heterogeneity in genetic regulations, for example GEs (gene expressions) by CNVs (copy number variations) and methylation, has been suggested but little investigated. Heterogeneity in genetic regulations can be linked with disease severity, progression, and other traits and is biologically important. However, the analysis can be very challenging with the high dimensionality of both sides of regulation as well as sparse and weak signals. In this article, we consider the scenario where subjects form unknown subgroups, and each subgroup has unique genetic regulation relationships. Further, such heterogeneity is ``guided" by a known biomarker. We develop an MSF (Multivariate Sparse Fusion) approach, which innovatively applies the penalized fusion technique to simultaneously determine the number and structure of subgroups and regulation relationships within each subgroup. An effective computational algorithm is developed, and extensive simulations are conducted. The analysis of heterogeneity in the GE-CNV regulations in melanoma and GE-methylation regulations in stomach cancer using the TCGA data leads to interesting findings.}


\maketitle

\section{Introduction}\label{sec1}

Heterogeneity is a hallmark of many complex diseases
\cite[]{mcclellan2010genetic,birbrair2019stem}. There are multiple ways of defining heterogeneity. In ``classic" biomedicine, heterogeneity is often defined based on clinical characteristics. With the increasing popularity of high-throughput profiling, there has been a surge of heterogeneity analysis based on high-dimensional molecular data
\cite[]{Sotiriou2003Breast,Burrell2013The}. A famous early example is the re-characterization of breast cancer subtypes based on gene expression (GE) profiles \cite[]{Sotiriou2003Breast}.
In more recent studies, integrated/multi-platform profiling has been extensively conducted, and such data
have been used to characterize breast \cite[]{shen2009integrative}, kidney \cite[]{cancer2016comprehensive}, and lung cancers \cite[]{chen2017multiplatform}. In this study, we take a strategy that differs significantly in most of the existing literature and examine disease heterogeneity in terms of genetic (and/or epigenetic) regulation patterns, under which subjects form subgroups, and different subgroups have unique genetic regulations, for example, of GEs by CNVs (copy number variations) and/or DNA methylation. Genetic regulations have been extensively studied in the literature \cite[]{bradner2017transcriptional,kagohara2018epigenetic}. The associations of genetic regulation relationships with disease severity, progression, and other traits have been noted. For example, Dutta et al. \cite[]{Dutta2011A} found that different subtypes of breast cancer had different GE-CNV regulation relationships, and such differences could be both quantitative (with the same signs but different magnitudes) and qualitative (with different signs). Similar findings were made in Safonov et al. \cite[]{SafonovImmune}. Differences in the GE-CNV regulations between cancer patients and normal controls have been identified for breast cancer \cite[]{Hyman2002Impact,Pollack2002Microarray}, colon cancer \cite[]{Platzer2002Silence}, and other cancers. Differences in the GE-microRNA regulations between cancer patients and normal controls have been also observed for multiple cancer types \cite[]{Liu2010Identifying,Lichtblau2017Comparative}. Such findings suggest that differences in genetic regulations may play important roles in  disease development and progression, and are critical in disease subtype characterization.  However, the aforementioned studies and those alike are all ``passive'', in the sense that they have predefined subject groups and then examine differences in genetic regulations.
In this study, we take a more ``proactive'' strategy. Specifically, our objective is to analyze genetic regulations for a set of subjects, who do not have labels, and identify previously unknown subgrouping structures under which subjects in each subgroup have unique genetic regulations. Such analysis has not been pursued in the literature and can provide a new way of defining disease heterogeneity and assist better understanding the biology of complex diseases. Here we also note that, in the literature, there have been quite a few advanced analyses on genetic regulations under the homogeneity assumption \cite[]{Wu2018}, some of which have demonstrated the advantage of penalization. As our main challenge is brought by the unknown heterogeneity, we will not further discuss these studies.

\begin{figure}[!tbp]
   \centering
   \includegraphics[scale=0.8]{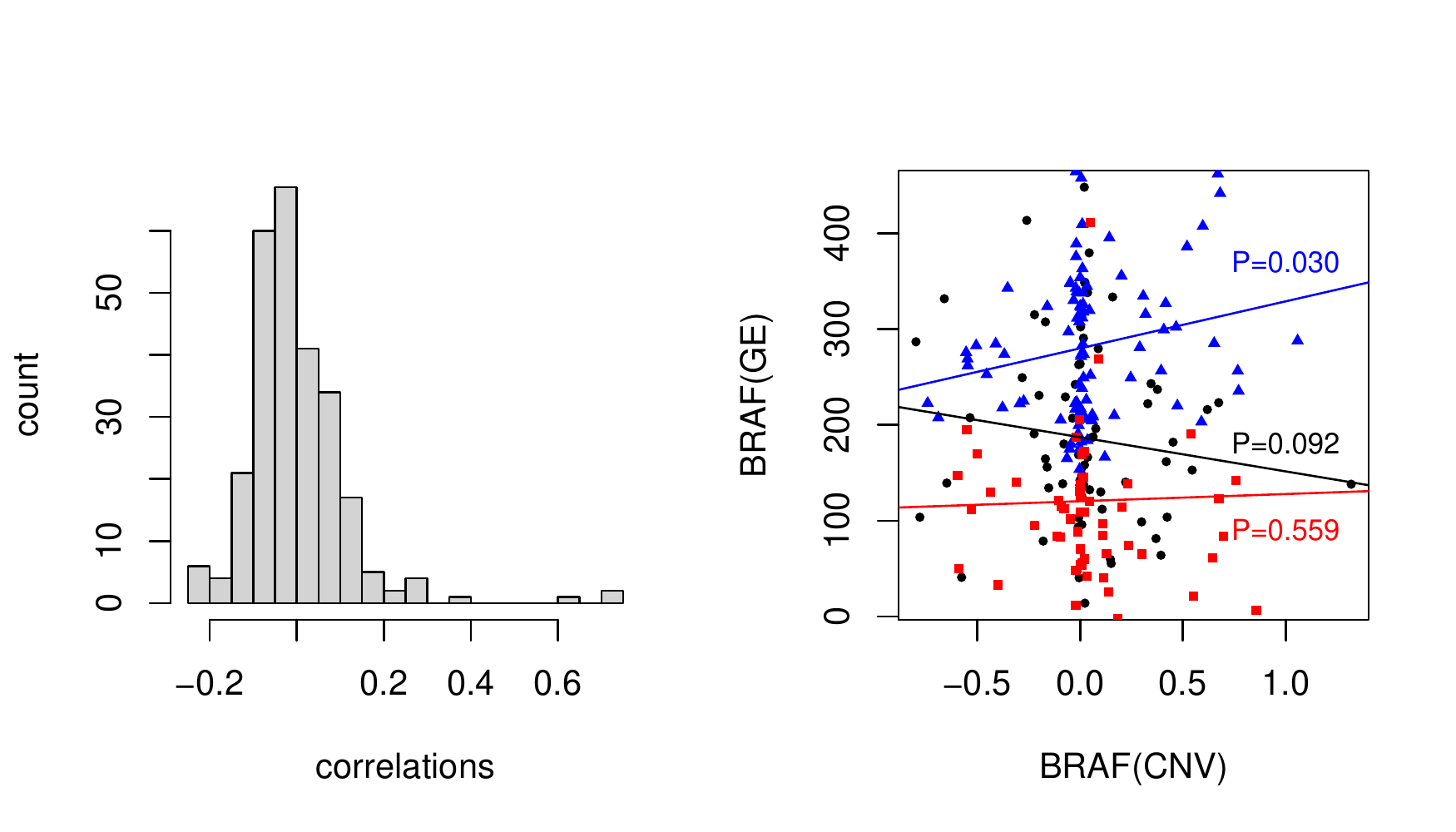}
   \caption{Analysis of TCGA melanoma data. Left: histogram of correlations between gene expression BRAF and 224 CNVs. Right: scatterplot of BRAF gene expression and CNV. Three colors correspond to three subgroups identified by the proposed method. }
   \label{fig:descriptive}
\end{figure}

To gain some insights into characteristics of the proposed analysis, we conduct a preliminary examination of the TCGA melanoma data (more details in Section \ref{sec:data}). In the left panel of Figure \ref{fig:descriptive}, we show the histogram of the correlations between the expression of gene BRAF and 224 CNVs. With the bulk of the correlations being small, it is reasonable to expect that one GE is only regulated by a small number of CNVs -- this is also supported by extensive genetic research \cite{Demichelis2012, Kumaran2018}. As such, quantifying genetic regulations poses a sparse variable selection and estimation problem. In the right panel of Figure 1, we examine the cis-acting regulation of BARF gene expression by CNV. The three colors correspond to the three subgroups identified by the proposed method, and the lines are generated using linear regression. {\color{blue} It is easy to see that, for different subgroups, the effects of CNV on the BARF gene are significantly different.}

\begin{figure}[htbp]
   \centering
    \includegraphics[scale=0.6]{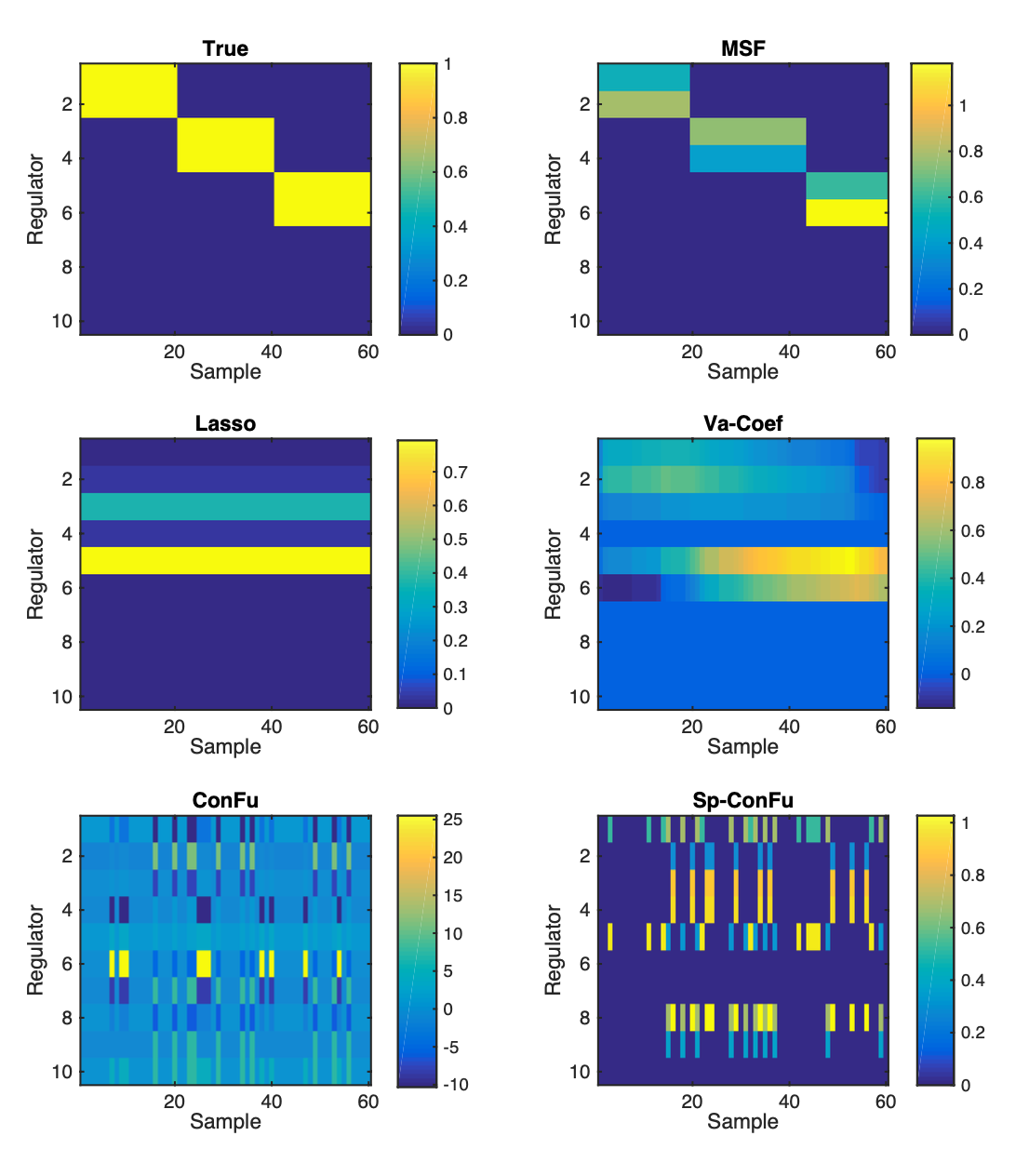} 
   \caption{Small example: heatmaps of true and estimated coefficients.}
   \label{fig:example}
\end{figure}

\noindent{\bf A small example}
We use a small simulation example to first examine the sparse variable selection/estimation aspect, demonstrate limitations of the existing approaches, and establish the demand for new methodological development. The heterogeneity/subgrouping aspect will be examined later in the article. A dataset with $n=60$ samples is simulated, and the sample sorting is guided by a biomarker (more details in Section 2). The values of $p=10$ regulators
(denoted as $x$'s) are generated from a multivariate normal distribution with an auto-regressive correlation structure where the correlation parameter $\rho=0.8$. The values of $q=10$ GEs (denoted as $y$'s) are generated from the following linear models with errors standard normally distributed:
\begin{equation*}
y_j^{(i)}=\left\{
\begin{array}{ll}
x_1^{(i)} + x_2^{(i)}+\epsilon_j^{(i)} & (i=1,\ldots,20),\\
x_3^{(i)} + x_4^{(i)}+\epsilon_j^{(i)} & (i=21,\ldots,40),\\
x_5^{(i)} + x_6^{(i)}+\epsilon_j^{(i)} &(i=41,\ldots,60).
\end{array}
\right.
\end{equation*}
Here the superscript is the subject index, and the subscript is the variable index. There are three subgroups with equal sizes, and each subgroup has a unique regulation model. The following approaches are applied for estimation and/or sparse variable selection:  (i) MSF (Multivariate Sparse Fusion), which is the approach developed in this study; (ii) Lasso,  which applies Lasso for sparse variable selection and estimation. Note that this approach does not have a mechanism to accommodate heterogeneity;  (iii) Va-Coef \cite[]{Hastie1993}, which is a varying-coefficient approach, assumes that each specific GE-regulator regulation coefficient is a smooth function of the underlying biomarker, estimates using cubic B-splines, and applies group Lasso penalization for variable selection. This approach can accommodate the differences in regulation coefficients across subjects;  (iv) ConFu \cite[]{Ma2016Estimating}, which estimates subgroup-specific coefficients via concave fusion but does not induce sparsity; and (v) Sp-ConFu, which advances from ConFu by applying Lasso to induce sparsity. Among the four alternative approaches, Lasso serves as a ``benchmark'', and the other three can accommodate heterogeneity and apply different popular strategies for estimation (and possibly selection). For a single replicate, we show as a demonstration the true and estimated coefficients of the 1st response variable in Figure \ref{fig:example}. For this specific response variable and others (results omitted), only the proposed approach correctly identifies the sparsity (and heterogeneity) structure. We also repeat simulation 100 times and examine variable selection performance using  TPR (true positive rate) and FPR (false positive rate), estimation performance using EMSE (estimation mean squared error), and prediction performance using PMSE (prediction mean squared error). Table \ref{tab:example} reports the summary statistics. 
It is noted that the drastically large EMSE of ConFu is caused by a lack of sparsity, which is effectively improved by Sp-ConFu. This small example suggests that, from the sparse variable selection and estimation perspectives, the existing approaches have unsatisfactory performance, and new development is needed.

The significance of understanding and quantifying heterogeneity for complex diseases has been well established in the literature. This study will be the first to model disease heterogeneity via genetic regulation. This perspective significantly differs from the existing ones \cite[]{shen2009integrative,cancer2016comprehensive,chen2017multiplatform}. As such, the proposed analysis can complement the existing literature, has the potential to leading to a deeper understanding of disease biology, and may also have clinical implications. For the proposed heterogeneity analysis, the small example above and other numerical studies in this article show that the existing approaches are not effective, and a new analysis approach is needed. The proposed approach is tailored to the genetic regulation-based heterogeneity analysis. It has some connections with but, more importantly,  significantly advances from the existing ones. Specifically, it advances from the existing genetic regulation studies by accommodating (unknown) heterogeneity. Compared to the finite mixture regression (FMR)-based approaches \cite[]{Shen2015Inference}, it can more conveniently determine the number of groups and in principle accommodate small groups. In addition, the existing FMR works are usually limited to a single response variable. The proposed approach also advances from the existing penalized fusion studies \cite[]{Ma2016Estimating, Yamada2017Localized} by having high-dimensional response variables and covariates. Moreover, this study may also provide a fresh look at the TCGA data. With advancements in paradigm, methodology, and data analysis, it is warranted beyond the existing literature.

 \begin{table}
 \centering
   \caption{Small example: summary statistics. In each cell, mean(standard deviation).}
   \label{tab:example}
    \setlength{\tabcolsep}{5pt}
    \renewcommand{\arraystretch}{1.1}
	\begin{tabular}{ccccc}
\toprule
   Method & \multicolumn{2}{c}{Variable selection} & EMSE & PMSE  \\
  \cmidrule{2-3}
  & TPR & FPR \\
  \midrule
MSF & 78\% (8\%) & 17\% (4\%) & 0.12 (0.02) & 1.45 (0.22)	\\
Lasso & 64\% (8\%) & 36\% (6\%) & 0.18 (0.01) & 1.99 (0.22)	\\
Va-Coef & 79\% (7\%) & 46\% (5\%)  &0.14 (0.01)	&1.32 (0.10)	\\
ConFu  & 100\% (0\%)  & 100\% (0\%) & 67.59 (39.02) & 1.17 (0.24)\\
Sp-ConFu  & 77\% (8\%)  & 42\% (6\%) & 0.23 (0.06) & 1.47 (0.26)\\
\bottomrule
\end{tabular}
\end{table}
\section{Methods}

\subsection{Data}
Consider a dataset with $n$ independent samples, each with $q$ response variables, and $p$ covariates. Here to fix ideas, we consider GEs as responses and their regulators (CNVs, methylation, microRNAs, etc.) as covariates. The proposed analysis is also applicable to other molecular measurements, for example proteins and their encoding gene expressions.
For the $i$th sample, let $Y^{(i)}=(y^{(i)}_1,y^{(i)}_2,\ldots,y^{(i)}_q)$ and $X^{(i)}=(x^{(i)}_1,x^{(i)}_2,\ldots,x^{(i)}_p)$ denote the vectors of response and covariate, respectively. Assume that the data have been standardized. Consider the multivariate regression model:
\begin{equation}
\label{eq:model}
	Y^{(i)}=X^{(i)}\bm B^{(i)}+\epsilon^{(i)},~i=1,2,\ldots,n,
\end{equation}
where $\epsilon^{(i)}$'s are random errors, and $\bm B^{(i)}=(\beta_{1}^{(i)},\beta_{2}^{(i)},\ldots, \beta_{q}^{(i)})$ is the $p\times q$ regression coefficient {\it matrix} with $\beta_j^{(i)}=(\beta_{j1}^{(i)},\beta_{j2}^{(i)},\ldots,
\beta_{jp}^{(i)})^\top$.

Further, we focus on the scenario where heterogeneity analysis is ``guided'' by a biomarker whose measurement is also available for the $n$ subjects. Specifically, consider the setting that the subjects have been sorted according to the biomarker's values. There are cutoffs of the biomarker, and subjects falling between two adjacent biomarker cutoffs share the same regulation model. It is noted that {\it the number and locations of cutoffs are unknown}. Studies referred to in Section 1 and in the literature suggest that heterogeneity in genetic regulation is often associated with disease severity and development. As such, in data analysis, we choose the biomarker as one with known association with clinical severity and development. Examples are provided in data analysis, and more discussions are provided below.

\noindent{\bf Remarks}
Linear regression is adopted for describing genetic regulations, which has been shown as effective in multiple studies \cite[]{Shi2015Deciphering, Peng2010Regularized}. To flexibly accommodate heterogeneity, the regression coefficient matrices are allowed to differ across subjects. This leads to a total of $p\times q\times n$ parameters to start with in modeling, which is much more challenging than many ``ordinary" high-dimensional problems.
Quantifying heterogeneity is equivalent to determining which subjects have the same regression coefficient matrices (and their values). We note that this differs significantly from the FMR and many other heterogeneity analysis works. In low-dimensional penalized fusion and other heterogeneity analysis, the guiding biomarker is not needed. However, we have briefly explored analysis without a guiding biomarker (and hence the samples are not ordered) and found that, with the high dimensionality of both $Y$ and $X$, penalized fusion is computationally intractable.
We note that this biomarker does not need to be precise. Consider for example subjects 1-4 (out of a sample of size $n$) which have the same genetic regulation model. Their biomarker A has two adjacent cutoffs 0.1 and 0.7, and values 0.2, 0.3, 0.4, and 0.5 for the four subjects. Biomarker B has two adjacent cutoffs -1.1 and 0.3, and values -0.9, -1.0, 0.0, and -0.2 for the four subjects. Although biomarkers A and B do not match, the proposed analysis can lead to the same heterogeneity structure with both biomarkers as guiding. Here we note that this small example is meant to demonstrate that the guiding biomarker requirement, although critical, is not stringent. In practice, the cutting off values are not assumed to be known {\it a priori}.

\subsection{Penalized estimation}

We propose the MSF estimate
\begin{equation}
\label{eq:estimate}
 \{\hat{\bm B}^{(i)}: i=1,2,\ldots, n\} = \hbox{argmin}\left\{\frac{1}{2}\sum\limits_{i=1}^{n}\lVert Y^{(i)}-X^{(i)}\bm B^{(i)}\rVert_2^2+ \lambda_1\sum\limits_{i=1}^{n}
\sum\limits_{j=1}^q\lVert\beta_{j}^{(i)}\rVert_1^2+\lambda_2
\sum\limits_{i=1}^{n-1}\sum\limits_{j=1}^{q}\lVert\beta_{j}^{(i)}
-\beta_{j}^{(i+1)}\rVert_2\right\},
\end{equation}
where $\lambda_1, \lambda_2\geq0$ are the tuning parameters. With the penalized estimates, a nonzero component corresponds to a genetic regulation. {\it Subjects $i$ and $l$ belong to the same subgroup (with the same genetic regulation model) if and only if $\hat{\bm B}^{(i)}=\hat{\bm B}^{(l)}$.} This approach is advantageous over the FMR and many others, as the number and structure of subgroups are automatically and simultaneously determined in the estimation process.

Overall, a penalized fusion 
strategy is taken. The first term of the objective function measures lack-of-fit. The first penalty term is mainly responsible for sparse estimation and variable selection. In particular, we adopt an $\ell_{12}$ penalty, which has also been referred to as the exclusive penalization \cite[]{Kong2014Exclusive}, to select a subset of relevant parameters within each $\beta_{j}^{(i)}$. Here $\beta_{j}^{(i)}$, which corresponds to the regulators of the $j$th response of subject $i$, is viewed as a group. The quadratic form of the penalty ensures that $\hat\beta_j^{(i)}\neq 0$, that is, all responses are regulated. This is sensible unless the measured regulator set is extremely ``biased", in which case it is not meaningful to conduct the genetic regulation-based analysis. It is noted that many alternative penalties, such as Lasso, cannot guarantee $\hat\beta_j^{(i)}\neq 0$. The second penalty takes a ``standard" fusion strategy and promotes zero differences (of regression coefficients) and hence grouping. It is noted that with the guiding biomarker, only differences between adjacent subjects are taken. Without the guiding biomarker, differences would need to be taken between any two subjects, leading to $n(n-1)/2$ terms. Our preliminary investigation suggests that when $p, q$ are not small, optimization with such an objective function is intractable. It is also noted that the proposed penalty combination leads to a convex optimization problem, which has a globally optimal solution and considerable advantages over concave penalties \cite[]{Ma2017A}. 

\noindent{\bf Remarks} As can be seen from (\ref{eq:estimate}), with different guiding biomarkers, the second penalty will be different. As described in Remarks in Section 2.1, if multiple biomarkers are sufficiently ``concordant'', they can lead to the same subgrouping and estimation results. Otherwise, different estimates are expected. In practice, we recommend selecting the guiding biomarker based on the specific biological contexts, as in our data analysis. Another possibility is to select based on statistical measures, for example prediction performance (in the same way as we compare with alternatives in numerical study). When the selected guiding biomarker aligns with differences in subtypes, the proposed approach can lead to meaningful subtype identification.

\subsection{Computation}
Let $\beta_j=\Big((\beta_j^{(1)})^\top,\ldots,(\beta_j^{(n)})^\top\Big)^\top$. The objective function in (\ref{eq:estimate}) is separable with respect to $\beta_j$ ($j=1,\ldots,q$), and thus the original optimization problem can be decomposed into $q$ subproblems:
\begin{equation}
\label{eq:subproblem}
\hbox{min}_{\beta_j}\frac{1}{2}\sum\limits_{i=1}^{n}\Big(y^{(i)}
-X^{(i)}\beta_j^{(i)}\Big)_2^2+\lambda_1\sum\limits_{i=1}^{n}\lVert\beta_{j}^{(i)}\rVert_1^2 +\lambda_2\sum\limits_{i=1}^{n-1}
\lVert\beta_{j}^{(i)}-\beta_{j}^{(i+1)}\rVert_2,
\end{equation}
which can be solved separately. Take the $j$th subproblem as an example. We first introduce a new set of parameters $\eta_j^{(i)}=\beta_j^{(i)}-\beta_j^{(i+1)}$. In addition, we also define a set of parameters $\delta_j^{(i)}=\beta_j^{(i)}$ to separate the $\ell_{12}$ penalization and the sum of squares operation imposed on $\beta_j^{(i)}$. Then optimization in (\ref{eq:subproblem}) can be reformulated as:
\begin{equation*}
\hbox{min}f(\beta_j,\eta_j,\delta_j)\equiv\frac{1}{2}
\sum\limits_{i=1}^n(y^{(i)}-X^{(i)}\beta_j^{(i)})^2+\lambda_1\sum\limits_{i=1}^n\lVert\delta_j^{(i)}\rVert_1^2+
\lambda_2\sum\limits_{i=1}^{n-1}\lVert\eta_j^{(i)}\rVert_2,
\end{equation*}
subject to
\begin{equation*}
\delta_j^{(i)}=\beta_j^{(i)},  i=1,\ldots,n, ~~
\eta_j^{(i)}=\beta_j^{(i)}-\beta_j^{(i+1)},  i=1,\ldots,n-1,
\end{equation*} $\eta_j=\Big((\eta_j^{(1)})^\top,\ldots,(\eta_j^{(n-1)})^\top\Big)^\top$ and  $\delta_j=\Big((\delta_j^{(1)})^\top,\ldots,(\delta_j^{(n)})^\top\Big)^\top$. To simplify notation, we omit the subscript $j$ in what follows. The augmented Lagrangian function reads
\begin{equation*}
\mathcal{L}(\beta,\eta,\delta,\theta,\gamma)  = f(\beta,\eta,\delta)+
\sum\limits_{i=1}^n(\theta^{(i)})^\top(\beta^{(i)}-\delta^{(i)})+\sum\limits_{i=1}^{n-1}(\gamma^{(i)})^\top(\beta^{(i)}-\beta^{(i+1)}-\eta^{(i)})+\frac{\rho}{2}\Big(\sum\limits_{i=1}^n\lVert\beta^{(i)}-\delta^{(i)}\rVert_2^2+\sum\limits_{i=1}^{n-1}\lVert\beta^{(i)}-\beta^{(i+1)}-\eta^{(i)}\rVert_2^2\Big).
\end{equation*}
Here  $\theta=\Big((\theta^{(1)})^\top,\ldots,(\theta^{(n)})^\top\Big)^\top$, $\gamma=\Big((\gamma^{(1)})^\top,\ldots,(\gamma^{(n-1)})^\top\Big)^\top$ are the Lagrange multipliers, and $\rho=2$ is the penalty parameter.
We iteratively update $\beta,\eta,\delta,\theta,\gamma$ using the ADMM technique. The update equations and algorithm are provided in the Appendix. The proposed approach involves two tuning parameters, which are selected using a grid search and the BIC criterion. More details are provided in the Appendix.

\noindent{\bf Realization} To facilitate data analysis within and beyond this study, we have developed a matlab code implementing the proposed approach and and made it publicly available at https://github.com/XnhuUcas/MSF. The proposed approach is computationally affordable. For example, with fixed tuning parameters, for a simulated dataset with $n=240$, $p=q=150$, and three subgroups, the analysis can be accomplished within 12 minutes using a laptop with standard configurations.The computation time can be greatly reduced by parallel computing. 

\vspace{5mm}

\section{Simulation}
\label{sec:simulation}
We set $p=q=150$ and $n=240$. The $p$ covariates are generated from a multivariate normal distribution with marginal means 0 and variances 1. We consider an auto-regressive (AR) correlation structure, where covariates $j$ and $k$ have correlation coefficient $\rho^{|j-k|}$ with $\rho=0.3$ and $0.8$. The $n$ samples belong to three subgroups. It is noted that the number of subgroups is not assumed to be known in analysis. We consider two subgroup structures: (a) balanced, where the three subgroups have the same sizes; and (b) unbalanced, where the three subgroups have sizes 60, 80, and 100. Denote $\bm \Omega_k$ ($k=1,2,3$) as the $p\times q$ coefficient matrix of the $k$th subgroup, for which we consider the following.

\noindent\underline{Unstructured}
$\bm \Omega_k$ ($k=1,2,3$) has six non-zero entries in each column. For the sparsity structure, we consider two scenarios: \\
(S1) Non-overlapping: 

\begin{equation*}
	\bm\Omega_1=
		\begin{pmatrix}
			 \Lambda       &\Lambda &\cdots & \Lambda   \\
			    0        &0&\cdots & 0  \\
			    0        &0&\cdots & 0  \\
 			 \vdots   &\vdots	    & \vdots & \vdots \\
			    0        &0 &\cdots & 0  \\
		\end{pmatrix},
	\bm\Omega_2=
		\begin{pmatrix}
			 \bm 0_6        & \bm 0_6             &\cdots & \bm 0_6  \\
			 \Lambda       &\Lambda &\cdots & \Lambda  \\
		     		0        & 0             &\cdots & 0  \\
 			 \vdots 	    & \vdots      & \vdots & \vdots \\
			   0        & 0             &\cdots & 0  \\
		\end{pmatrix},
	\bm\Omega_3=
		\begin{pmatrix}
			 \bm 0_{6}        & \bm 0_{6}     &\cdots & \bm 0_{6}  \\
			 \bm 0_6        & \bm 0_6             &\cdots & \bm 0_6  \\
			 \Lambda       &\Lambda &\cdots & \Lambda \\
 			 \vdots 	    & \vdots      & \vdots & \vdots \\
			   0        & 0             &\cdots & 0  \\
		\end{pmatrix},
\end{equation*}

(S2) Overlapping:
{
\begin{equation*}
	\bm\Omega_1=
		\begin{pmatrix}
			    \Lambda_1     & \Lambda_1       &\cdots & \Lambda_1   \\
			    0        & 0             &\cdots & 0  \\
			    0        & 0             &\cdots & 0  \\
 			 \vdots 	    & \vdots      & \vdots & \vdots \\
			    0        & 0             &\cdots & 0  \\
		\end{pmatrix},
	\bm \Omega_2=
		\begin{pmatrix}
			 \bm 0_2        & \bm 0_2             &\cdots & \bm 0_2  \\
			 \Lambda_2     & \Lambda_2       &\cdots & \Lambda_2   \\
		     		0        & 0             &\cdots & 0  \\
 			 \vdots 	    & \vdots      & \vdots & \vdots \\
			   0        & 0             &\cdots & 0  \\
		\end{pmatrix},
	\bm \Omega_3=
		\begin{pmatrix}
			 \bm 0_{5}        & \bm 0_{5}     &\cdots & \bm 0_{5}  \\
			   \Lambda_3     & \Lambda_3       &\cdots & \Lambda_3   \\
		     		0        & 0             &\cdots & 0  \\
 			 \vdots 	    & \vdots      & \vdots & \vdots \\
			   0        & 0             &\cdots & 0  \\
		\end{pmatrix}.
\end{equation*}
}
Here $\bm 0_{s}$ is a $s\times 1$ zero vector, $\Lambda$ and $\Lambda_k$ $(k=1,2,3)$ are $6\times 1$ vectors. Each entry of $\Lambda$ is 1. $\Lambda_k$ $(k=1,2,3)$ have entries independently drawn from \emph{Unif}$(-2.2,-2)$, \emph{Unif}$(1,1.2)$, and \emph{Unif}$(2.5,2.8)$, respectively. The response variables are generated via the linear regression model (\ref{eq:model}), in which the random errors are independently generated from normal distributions with mean 0 and variance
$\sigma^2=1$ and 3.

\noindent\underline{Block structure}. $\bm\Omega_k$ has a block diagonal structure with $30$ blocks. For the sparsity structure, consider two scenarios: \\
(S3) Non-overlapping: 

{
\begin{equation*}
	\bm\Omega_1=
		\begin{pmatrix}
			  \Gamma\circ K_1     &       &\\
			           &  \ddots   &  \\
			          &       & \Gamma \circ K_{10} \\
		\end{pmatrix},
            \bm\Omega_2=
		\begin{pmatrix}
			           \Gamma \circ K_{11}  &     &        \\
			          &   \ddots      &         \\
				      &      &      \Gamma \circ K_{20}  \\
		\end{pmatrix},
	 \bm\Omega_3=
		\begin{pmatrix}
			           &    \Gamma \circ K_{21}      &        &      \\
				      &      &    \ddots &\\
				     &        &        &     \Gamma \circ K_{30}   \\
		\end{pmatrix}.
\end{equation*}
}

(S4) Overlapping: 
{
\begin{equation*}
	\bm\Omega_1=
		\begin{pmatrix}
			  \Gamma_1\circ K_1     &            \\
			          &  \ddots   &     \\
				      &        & \Gamma_1\circ K_{30}
		\end{pmatrix},
            \bm\Omega_2=
		\begin{pmatrix}
			  \Gamma_2 \circ K_1     &            \\
			          &  \ddots   &     \\
				      &        & \Gamma_2 \circ K_{30}
		\end{pmatrix},
	 \bm\Omega_3=
		\begin{pmatrix}
			  \Gamma_3 \circ K_1     &            \\
			          &  \ddots   &     \\
				      &        & \Gamma_3 \circ K_{30}
		\end{pmatrix}.
\end{equation*}
}

Here $\circ$ denotes the entry-wise product. $\Gamma$, $\Gamma_k$ ($k=1,2,3$), and $K_t$ $(t=1,2,\ldots,30)$ are $5\times 5$ matrices. The entries of $K_t$ are independently generated from a Bernoulli distribution with a success probability of 0.8. Each entry of $\Gamma$ is 1. $\Gamma_k$ has entries independently drawn from \emph{Unif}$(-2.2,-2)$, \emph{Unif}$(1,1.2)$, and \emph{Unif}$(2.5,2.8)$ for $k=1,2,3$, respectively. We generate the response variables from the multivariate regression model (\ref{eq:model}). The random errors are generated independently from $N(\bm 0,0.5^2\times\hbox{AR}(\phi))$ with $\phi=0.3$ and $0.8$. The dependent random errors reflect the fact that there may be additional regulators that are not measured.

\renewcommand\arraystretch{1.0}
\begin{table}
 \caption{Simulation under setting S1. In each cell, mean(sd).}
\label{sim:S1}
 \begin{center}
    \setlength{\tabcolsep}{5pt}
    \renewcommand{\arraystretch}{1.1}
	\begin{tabular}{ccccccccccc}
\hline
    &  &  & & \multicolumn{2}{c}{Heterogeneity identification} & & \multicolumn{2}{c}{Variable selection} \\
	\cline{5-6} \cline{8-9}
   Structure & \multicolumn{2}{c}{($\rho$,$ \sigma^2$)} & Method &  Num	& Rand	   && TPR	    &FPR 	& EMSE & PMSE  \\\hline
\multirow{16}{*}{Balanced} &
\multicolumn{2}{c}{\multirow{4}{*}{(0.3,1)}}
&	    MSF         &	3.02(0.14)     &	0.971(0.024)    &&	97.3\%(2.6\%)	&	1.4\%(0.3\%)	& 	0.006(0.002)	&	1.709(0.474)	\\
& & &	RespClust	&	3.38(0.73)	&	0.803(0.086)  	&&	72\%(12\%)	    &	1.9\%(0.8\%)	&	0.02(0.006)	    &	4.875(1.445)	\\
& & &	ResiClust	&	3.40(0.70) 	&	0.806(0.062)	&&	72.7\%(11.6\%)	&	1.9\%(0.9\%)	&	0.02(0.005)	    &	4.813(1.245)	\\
& & &	RiFuClust 	&	2.48(0.50) 	&	0.835(0.097) 	&&	84.6\%(11.8\%)	&	2.2\%(1.1\%)	&	0.017(0.009)	&	3.981(1.761) 	\\
\cline{2-11}
& \multicolumn{2}{c}{\multirow{4}{*}{(0.3,3)}}
&	    MSF	        &	3.04(0.20) 	&	0.959(0.032) 	&&	93.3\%(4.8\%)	&	1.7\%(0.3\%)	&	0.012(0.003)	&	4.128(0.546) 	\\
& & &	RespClust	&	3.28(0.73) 	&	0.811(0.073) 	&&	62.7\%(12.9\%)	&	1.6\%(0.6\%)	&	0.025(0.005)	&	7.335(1.254)	\\
& & &	ResiClust	&	3.22(0.71) 	&	0.811(0.061) 	&&	63.2\%(11.1\%)	&	1.7\%(0.7\%)	&	0.025(0.004)	&	7.357(1.087) 	\\
& & &   RiFuClust 	&	2.34(0.48) 	&	0.801(0.08) 	&&	70.3\%(13.7\%)	&	2.1\%(0.8\%)	&	0.024(0.006)	&	7.13(1.506)		\\
\cline{2-11}
& \multicolumn{2}{c}{\multirow{4}{*}{(0.8,1)}}
&	    MSF	        &	3.00(0.00) 	&	0.982(0.012) 	&&	97.7\%(1.9\%)	&	0.5\%(0.1\%)	&	0.006(0.002)	&	1.753(0.599)	\\
& & &	RespClust	&	3.32(0.62) 	&	0.833(0.071) 	&&	81.1\%(8.7\%)	&	1.5\%(0.7\%)	&	0.018(0.006)	&	4.786(1.455)	\\
& & &	ResiClust	&	3.32(0.79) 	&	0.811(0.078) 	&&	80.2\%(8.3\%)	&	1.7\%(0.8\%)	&	0.02(0.006)	    &	5.014(1.883)	\\
& & &	RiFuClust 	&	2.96(0.20) 	&	0.946(0.057)	&&	94.3\%(5.3\%)	&	0.7\%(0.5\%)	&	0.009(0.005)	&	2.468(1.284)	\\
\cline{2-11}
& \multicolumn{2}{c}{\multirow{4}{*}{(0.8,3)}}
&	    MSF         &	3.02(0.14) 	&	0.976(0.017) 	&&	93.5\%(2.6\%)	&	0.6\%(0.1\%)	&	    0.011(0.002)	&	3.896(0.452) 	\\
& & &	RespClust	&	3.46(0.68) 	&	0.803(0.059) 	&&	72.7\%(6.6\%)	&	1.5\%(0.6\%)	&	0.024(0.004)	&	7.588(1.337) 	\\
& & &	ResiClust	&	3.52(0.54) 	&	0.823(0.064)	&&	74.6\%(7.7\%)	&	1.3\%(0.5\%)	&	0.022(0.005)	&	7.161(1.386)	\\
& & &	RiFuClust 	&	2.86(0.35) 	&	0.928(0.077) 	&&	88.9\%(7.6\%)	&	0.9\%(0.7\%)	&	0.015(0.006)	&	4.909(1.643)	\\\hline
\multirow{16}{*}{Unbalanced} &
\multicolumn{2}{c}{\multirow{4}{*}{(0.3,1)}}
&	    MSF	        &	3.00(0.00) 	&	0.962(0.027) 	&&	96.8\%(2.3\%)	&	1.5\%(0.3\%)	&	0.006(0.003)	&	1.696(0.451) 	\\
& & &	RespClust	&	3.42(0.73) 	&	0.797(0.066) 	&&	72.6\%(12.7\%)	&	2\%(0.7\%)	    &	0.02(0.006)	    &	4.55(1.208) 	\\
& & &	ResiClust	&	3.42(0.73) 	&	0.799(0.066) 	&&	72.9\%(12.8\%)	&	1.9\%(0.8\%)	&	0.02(0.006)	    &  	4.531(1.216)	\\
& & &	RiFuClust 	&	2.38(0.49) 	&	0.84(0.077) 	&&	83.5\%(11\%)	    &	2.4\%(1.1\%)	&	0.017(0.007)	&	3.926(1.468)	\\
\cline{2-11}
& \multicolumn{2}{c}{\multirow{4}{*}{(0.3,3)}}
&	    MSF	        &	3.06(0.24) 	&	0.961(0.032) 	&&	93.4\%(5.7\%)	&	1.6\%(0.3\%)	&	0.011(0.003)	&	4.015(0.501)	\\
& & &	RespClust	&	3.40(0.64) 	&	0.798(0.068)    &&	60.2\%(14.8\%)	&	1.4\%(0.5\%)	&	0.025(0.005)	&	7.232(1.278) 	\\
& & &   ResiClust	&	3.36(0.63) 	&	0.803(0.063) 	&&	61.1\%(13.9\%)	&	1.4\%(0.6\%)	&	0.025(0.005)	&	7.185(1.173)	\\
& & &	RiFuClust 	&	2.22(0.42) 	&	0.816(0.058) 	&&	70.3\%(13.7\%)	&	2\%(0.6\%)	    &	0.023(0.005)	&	7.115(1.322) 	\\
\cline{2-11}
& \multicolumn{2}{c}{\multirow{4}{*}{(0.8,1)}}
&	    MSF	        &	3.00(0.00) 	&	0.977(0.015) 	&&	97.4\%(2.2\%)	&	0.5\%(0.1\%)	&	0.006(0.002)	&	1.738(0.549) 	\\
& & &   RespClust   &	3.50(0.71) 	&	0.812(0.068) 	&&	81.1\%(8\%)	    &	1.5\%(0.6\%)	&	0.019(0.006)	&	4.667(1.691)	\\
& & &	ResiClust	&	3.30(0.81) 	&	0.797(0.069) 	&&	80.4\%(7\%)	    &	1.7\%(0.7\%)	&	0.019(0.005)	&	5.002(1.639) 	\\
& & &	RiFuClust 	&	2.86(0.35) 	&	0.932(0.063) 	&&	93\%(6.4\%)	    &	0.9\%(0.6\%)	&	0.01(0.006)	    &	2.569(1.445)	\\
\cline{2-11}
& \multicolumn{2}{c}{\multirow{4}{*}{(0.8,3)}}
&	    MSF	        &	3.04(0.20) 	&	0.975(0.023) 	&&	93.5\%(2.5\%)	&	0.7\%(0.1\%)	&	0.011(0.002)	&	3.898(0.503)	\\
& & &   RespClust	&	3.40(0.61) 	&	0.81(0.077) 	&&	75.2\%(7.5\%)	&	1.5\%(0.6\%)	&	0.023(0.005)	&	7.082(1.754) 	\\
& & &   ResiClust	&	3.22(0.65) 	&	0.797(0.088) 	&&	73.2\%(7.6\%)	&	1.6\%(0.6\%)	&	0.024(0.005)	&	7.521(1.624) 	\\
& & &	RiFuClust 	&	2.84(0.37) 	&	0.926(0.065) 	&&	87.9\%(7.2\%)	&	0.9\%(0.5\%)	&	0.015(0.005)	&	4.991(1.533) 	\\\hline
 	\end{tabular}
 	 \end{center}
\end{table}



The example in Section 1 has demonstrated the ineffectiveness of the alternatives in sparse selection and estimation. Hence here  we focus on comparing with the following approaches which emphasize more on the subgrouping (clustering) aspect: (a) Response-based clustering (RespClust), which first clusters the samples based on the $q$ response variables, and then applies Lasso to each subgroup; (b) Residual-based clustering (ResiClust), which applies Lasso under the homogeneity assumption $Y^{(i)}=X^{(i)}\bm B+\epsilon^{(i)}$, groups samples based on the residuals $Y^{(i)}-X^{(i)}\hat{\bm B} $, and then applies Lasso again to each subgroup; and (c) Ridge fusion clustering (RiFuClust), which modifies the proposed approach by revising the second penalty to a ridge type (which shrinks differences in coefficients but not to exactly zero), groups samples based on $\hat{\bm B}^{(i)}$'s, and then applies Lasso to each subgroup. 
For the sake of fairness, information from the guiding biomarker is also incorporated in grouping with the alternative approaches. More details are provided in the Appendix. To the best of our knowledge, there is no existing approach in the literature that conducts exactly the proposed analysis. The above alternative approaches share certain common ground with the proposed. As such, the comparison is sensible. We have briefly experimented with FMRs and encountered methodological and computational difficulties caused by the high dimensionality of responses and covariates.

In evaluation, we are interested in both the identification of heterogeneity (i.e., subgrouping performance) and variable selection accuracy (as all approaches generate sparse estimation). For the evaluation of heterogeneity identification, we consider two measures, namely the estimated number of subgroups (Num) and Rand Index (Rand)\cite[]{rand1971}, where Rand Index measures the agreement between the structure of the estimated subgroups and that of the true subgroups. We use TPR and FPR to evaluate variable selection performance. Here the definitions are the same as in the literature. In addition, we also evaluate estimation and prediction performance. Specifically, estimation performance is evaluated using EMSE (estimation MSE), which is defined as $\sum_{i=1}^n\parallel \hat{\bm B}^{(i)}-\bm B^{(i)}\parallel_F^2/npq$. Prediction performance is evaluated using PMSE (prediction MSE), which is defined as $\sum_{i=1}^n\parallel \hat{Y}^{(i)}-Y^{(i)}\parallel_2^2/nq$. 

With the proposed MSF approach, we first examine the values of BIC as a function of $\lambda_1$ and $\lambda_2$. Figure \ref{fig:BIC} in the Appendix presents the BIC function for a random replicate under setting S1, balanced structure, and $(\rho,\sigma^2)=(0.3,1)$. The optimal point with $(\lambda_1,\lambda_2)=(0.1,140)$ is clearly identified. We have also examined a few other replicates and observed similar patterns. We then compute summary statistics based on 200 replicates. Results under setting S1 are reported in Table \ref{sim:S1}, and those under settings S2-S4 are reported in Tables \ref{sim:S2}-\ref{sim:S4} in the Appendix. It is observed that across the whole simulation spectrum, the proposed approach has favorable performance in the identification of heterogeneity structure. Consider for example setting S1, balanced structure, and $(\rho, \sigma^2)=(0.3, 3)$. The mean values of the estimated number of subgroups are 3.04 (MSF), 3.28 (RespClust), 3.22 (ResiClust), and 2.34 (RifuClust). MSF has the value closest to 3, the true number of subgroups. The four approaches have Rand Index values 0.959 (MSF), 0.811 (RespClust), 0.811 (ResiClust), and 0.801 (RiFuClust). As another example, consider setting S3, unbalanced structure, and $(\rho, \phi)=(0.3, 0.3)$ (Table \ref{sim:S3}). For the identification of heterogeneity structure, the four approaches have number values 3.02 (MSF), 4.34 (RespClust), 4.42 (ResiClust), and 5 (RifuClust), and Rand Index values 0.983 (MSF), 0.822 (RespClust), 0.822 (ResiClust), and 0.848 (RiFuClust). The proposed approach is also superior in variable selection. Consider for example setting S2, unbalanced structure, and $(\rho, \sigma^2)=(0.8, 1)$ (Table \ref{sim:S2}). The (TPR, FPR) values are (0.904, 0.006) for MSF, (0.784, 0.008) for RespClust, (0.773, 0.008) for ResiClust, and (0.935, 0.009) for RiFuClust. For an alternative evaluation of variable selection performance, we consider a sequence of tunings, and generate ROC curves for all methods (Figure \ref{fig:roc}). The results re-confirm that the proposed approach has advantages in variable selection. Favorable estimation and prediction performance is also observed. Consider for example setting S1, balanced structure, and $(\rho, \sigma^2)=(0.3,1)$ (Table 2). The (EMSE, PMSE) values are (0.006, 1.709) for MSF, (0.020, 4.875) for RespClust, (0.02, 4.813) for ResiClust, and (0.017, 3.981) for RiFuClust. 

We conduct additional simulation to examine dependence of the proposed approach on $\Gamma_k$. Specifically, the entries in $\Gamma_k$ are independently drawn from truncated normal distributions with ranges $(-2.2,-2)$, $(1,1.2)$, and $(2.5,2.8)$ for $k=1,2,3$, respectively. The other settings are the same as described above. We use setting S4 as a representative and present the results in Table \ref{sim:S4_2} (Appendix).
The proposed approach is again observed to have favorable performance in heterogeneity identification, variable selection, estimation, and prediction.

\section{Data analysis}
\label{sec:data}

TCGA is a collective effort organized by the NCI/NIH, under which data on multiple types of molecular measurements have been collected on the same subjects for multiple cancer types. 
Several heterogeneity analysis studies have been conducted on the TCGA data \cite[]{
Sun2018Identification,Sun2019An}, although it is noted that they have taken strategies significantly different from the proposed. Data analyzed here are downloaded from the cBioPortal website (www.cbioportal.org). Standard data processing is conducted following the literature \cite[]{Dewey2011RSEM}. 

\subsection{Melanoma data}
In the analysis of melanoma data, we follow the literature \cite[]{Jiang2016Integrated} and focus on white patients who had non-glabrous skin and neo-adjuvant therapy. To quantify heterogeneity, we analyze the regulations of GEs by CNVs. In the original data, a total of 20, 531 RNAseq gene expression measurements and 24,776 CNVs are available. We consider the processed level-3 gene expression data, and refer to the literature \cite[]{Dewey2011RSEM} for detailed information on data generation and processing. In principle, there is no limit on the numbers of response and covariate variables that can be analyzed. However, whole-genome analysis can bring unstable estimation and tremendous computational challenges, especially considering the limited sample size. With the additional consideration that genetic regulations are often ``localized", we conduct a pathway-based analysis. This strategy has also been taken in multiple published studies \cite[]{Shi2015Deciphering}. Specifically, the KEGG pathway information (c2.cp.kegg.v6.2.symbols.gmt) is retrieved from the Broad Institute's website. We focus on the MAPK signaling pathway, which has been suggested as playing a critical role in melanoma \cite[]{Hidetoshi2006The}. This pathway has also been analyzed in published genetic regulation studies \cite[]{kunz2019modelling}. The final analyzed data contains 214 GE and 264 CNV measurements on 224 subjects. For the guiding biomarker, we use Breslow thickness, which is a confirmed biomarker for prognosis. The median Breslow thickness is 2 (range 0-50).

\begin{figure}[htbp]
   \centering
   \includegraphics[scale=0.15]{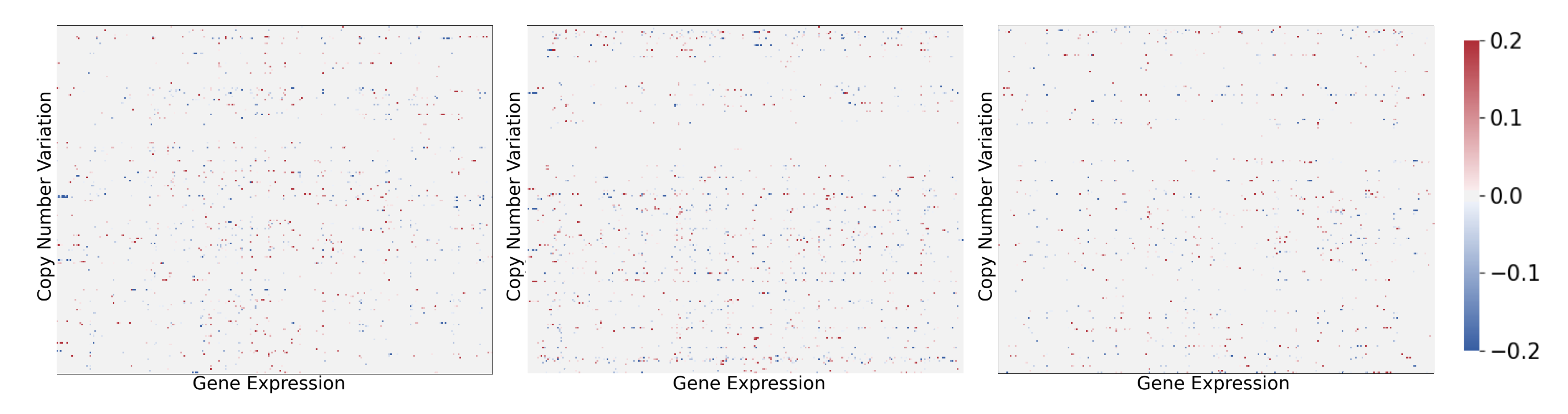}
   \caption{Analysis of melanoma data using the proposed approach: heatmaps of the estimated regulation coefficient matrices for the three subgroups. {\color{blue}The three matrices are significantly different ($p$-values$<0.001$).}}
   \label{fig:melanoma.heat}
\end{figure}

\begin{table}
\caption{Analysis of melanoma data using the proposed approach: regression coefficients for gene expression BRAF.}
\label{tab:BRAF}
 \begin{center}
    \setlength{\tabcolsep}{2.5pt}
    \renewcommand{\arraystretch}{1.0}
	\begin{tabular}{lccc} \hline
 & Subgroup 1 & Subgroup 2 & Subgroup 3 \\ \hline
BRAF & 0.228 & 0.365 &  \\
CACNA1D &  &  & 0.036 \\
CACNA1H & -0.056 &  &  \\
CACNA1S & -0.035 &  &  \\
PRKACG & -0.007 &  &  \\
PTPN7 & -0.004 &  &  \\
RASGRF1 &  & 0.185 &  \\
NRAS &  & 0.174 &  \\
FGF3 &  & -0.165 &  \\
MAPK3 &  & -0.151 &  \\
MAP3K7 &  & 0.118 &  \\
MAP2K6 &  &  & -0.157 \\
MAP2K7 &  & -0.118 &  \\
NGF &  & 0.113 &  \\
PTPRR &  & -0.110 &  \\
MRAS &  & 0.094 &  \\
FLNC &  &  & 0.136 \\
DUSP7 &  &  & 0.096 \\
RPS6KA6 &  &  & -0.022 \\
MECOM &  &  & 0.015 \\
\hline
 	\end{tabular}
\end{center}
\end{table}

With the proposed approach, we select the tuning parameters by minimizing the BIC, in the same way as in simulation. Results are briefly presented in Figure \ref{fig:BIC_data} in the Appendix. Under the optimal tunings, three subgroups, with with sizes 73, 93, and 58, are identified. 
The heatmaps of the three regulation coefficient matrices are shown in Figure \ref{fig:melanoma.heat}. We test the differences between these matrices using the generalized component approach \cite[]{Gregory2015A}, and the resulted $p$-values are highly significant ($<$0.001). More detailed subgrouping and estimation results are provided in the Supplementary Materials. We consider BRAF, a well-established melanoma genetic marker, as an example, and show its estimated regulation coefficients in Table \ref{tab:BRAF}. Significant differences across the subgroups are observed. In particular, except for the cis-acting CNV BRAF, the trans-acting regulating CNVs for the three subgroups have no overlapping. CNV BRAF has the largest and positive coefficients for the first two subgroups, and such a result is ``as expected". It is interesting to observe that it is not regulating for the third subgroup. As discussed in Section 1 and published literature, heterogeneity in genetic regulation is oftentimes associated with disease severity and development. However, biologically and statistically, there is no guarantee that such associations exist. We conduct an exploratory analysis and compare clinical features of samples in different subgroups and find that multiple features, such as tumor stage and age, differ significantly. In addition, prognosis also differs significantly across subgroups, which is expected as Breslow thickness is the guiding biomarker.

Analysis is also conducted using the alternatives, which all identify five subgroups. Detailed subgrouping and estimation results are available in the Supplementary Materials. Comparing the subgrouping structures of the proposed and alternative approaches reveals significant differences ($p$-values are shown in Table \ref{tab:pval} in the Appendix).  The variable selection results are compared in Table \ref{tab:compare} (Appendix), where we also observe significant differences. As has been noted in many published studies, it is difficult to objectively assess which set of heterogeneity analysis result is more sensible. We conduct the following evaluations, which may provide support to a certain extent. We first evaluate prediction performance. For a given subgrouping structure, subjects within each subgroup are partitioned into a training and a testing set with sizes 3:1. Estimation is conducted using the training set, and prediction is made with the testing test. With 100 random partitions, the average PMSEs are calculated as 0.918 (MSF), 1.050 (RespClust), 1.036 (ResiClust), and 1.003 (RifuClust). Second, we examine stability of the four approaches, in terms of subgroup identification and variable selection. Specifically, for subgroup identification, 10\% of the subjects are randomly removed from the original datasets. Let $\bm G=(g_{il})_{n\times n}$ be the adjacency matrices, where the $(i,l)$th element $g_{il}=1$ if subjects $i, l$ belong to the same subgroup and 0 otherwise. Denote $\hat{\bm G}$ and ${\bm G}^{(T)}$ as the adjacency matrices of the estimated and true subgroups, respectively. In data analysis, the ``truth'' is taken as the estimate using all subjects. Then we define the stability measure as $M_{sta}=\frac{1}{n^2}\sum_{il}|\bm{\hat G}_{il}-\bm {G}^{(T)}_{il}|$, following the definition proposed in the literature \cite[]{Teran2017Assisted}. The mean values of $M_{sta}$'s over 100 random replicates are 0.038 (MSF), 0.067 (ResiClust), 0.106 (RifuClust), and 0.082 (RifuClust). The proposed method has the highest stability in subgroup identification. We use the Observed Occurrence Index (OOI) to evaluate the stability of variable selection. The mean OOI values are 0.810 (MSF), 0.747 (RespClust), 0.751 (ResiClust), and 0.687 (RiFuClust). With the subgrouping structures identified by the alternatives, we also compare clinical features across subgroups and find less significant differences. Overall, the proposed approach seems to have more sensible findings.

\subsection{Stomach cancer data}
To demonstrate the broad applicability of the proposed analysis, here we consider GE and DNA methylation, an epigenetic regulating mechanism. It is again noted that the proposed analysis/approach does not demand the collection of all relevant regulators. 
In the original data, a total of 20, 531 gene expression measurements and 16, 764 methylation measurements are available. The DNA methylation data are log2 transformed. As in the last subsection, we focus on the MAPK signaling pathway. The final analyzed data contain 214 GEs and 229 methylation measurements. With the consideration that DNA methylation patterns and stomach cancer development/progression both heavily depend on age, we choose age as the guiding biomarker. In the raw dataset, there are 438 patients. After removing records with missing age, GE, and methylation measurements, a total of 367 patients are included in the final analysis. The median age is 67 (range 30-90).

The proposed approach identifies four subgroups, with sizes 99, 70, 96, and 102. The heatmaps of the four regulation coefficient matrices are shown in Figure \ref{fig:stomach.heat} (Appendix). The tests of differences between these matrices again lead to highly significant $p$-values ($<$0.001). More detailed subgrouping and estimation results are available in the Supplementary Materials. The regression coefficients for a representative example, gene TGFBR2, are shown in Table \ref{tab:stomach_coef} (Appendix). Methylation TGFBR2 has regulating effects in all four subgroups. Otherwise, the four subgroups do not have overlapping regulating methylations. In our exploratory analysis, we find certain important features, such as mutation burden, differ significantly across the four subgroups.

Analysis is also conducted using the alternatives. ResiClust and RespClust identify five subgroups, and RifuClust identify six subgroups. The differences in subgrouping results between the proposed and alternative approaches are highly significant (Table \ref{tab:pval} in the Appendix). The comparison of variable selection results in Table \ref{tab:compare} (Appendix) also suggests significant differences. In prediction evaluation, the average PMSE values are 0.902 (MSF), 0.948 (RespClust), 0.945 (ResiClust), and 0.945 (RiFuClust). In stability evaluation, the average $M_{sta}$ values are 0.041 (MSF), 0.055 (ResiClust), 0.045 (RifuClust), and 0.126 (RifuClust). The OOI values are 0.985 (MSF), 0.949 (RespClust), 0.943 (ResiClust), and 0.895 (RiFuClust). Comparing clinical features across subgroups for the alternatives leads to insignificant differences. Overall, it again seems that the proposed heterogeneity analysis is more sensible.

\section{Discussion}
Heterogeneity has important implications for many complex diseases. In this study, we have taken a strategy significantly different from those in the literature and quantified heterogeneity based on genetic regulations, which, as suggested by published studies, can be related to disease severity and development and have important practical implications. We have developed a novel new analysis approach. It takes the penalized fusion strategy, which has been recently developed and has multiple advantages over the FMR and other heterogeneity analysis strategies. It also advances from the existing penalized fusion studies by having both high-dimensional responses and covariates, by inducing sparsity, and by taking advantage of the guiding biomarker. Simulation has demonstrated satisfactory practical performance, and in data analysis, it has led to sensible findings significantly different from the alternatives.

There have been multiple different ways of defining heterogeneity. It is noted that heterogeneity defined based on molecular measurements is not necessarily associated with subtype or clinical traits. In addition, it remains unclear how to generate consensus from different heterogeneity definitions. In our description, we have mostly used gene expressions and their regulators as an example. Similar analysis can be conducted, for example, using the (proteins, gene expressions) dual. The penalized fusion technique has great flexibility and other advantages not shared by the FMR and other heterogeneity analysis techniques. However, it becomes computationally intractable when both responses and covariates are high-dimensional. This is resolved by introducing the guiding biomarker, which has biologically sensible interpretations. We defer to future research to develop techniques that can accommodate the absence of the guiding biomarker. In the literature \cite[]{Wu2018}, there are alternative, possibly more complicated, models for genetic regulations. We conjecture that it may be possible to combine such models with the proposed heterogeneity analysis via fusion. As considerable new developments will be needed, we postpone exploring this possibility to future research. In data analysis, heterogeneity has been identified for melanoma and stomach cancers. We defer to future research for biological interpretations.

\section*{Acknowledgments}
We thank the editor and reviewers for their careful review and insightful comments, which have led to a significant improvement of the article.
This study was partly supported by the Bureau of Statistics of China (2019LZ11), Fund for building world-class universities (disciplines) of Renmin University of China, NIH (CA216017, CA241699), NSF (1916251), and a Yale Cancer Center Pilot Award. The computer resources were provided by Public Computing Cloud Platform of Renmin University of China.\vspace*{-12pt}


\begin{thebibliography}{10}
\providecommand \doibase [0]{http://dx.doi.org/}%

\bibitem{mcclellan2010genetic}
McClellan J, King M. Genetic heterogeneity in human disease. {\it Cell.}
  2010\string; 141(2)\string: 210-217.

\bibitem{birbrair2019stem}
Birbrair A. {\it Stem Cells Heterogeneity in Different Organs}.
\newblock Berlin Heidelberg: Springer.
\newblock 2019.

\bibitem{Sotiriou2003Breast}
Sotiriou C, Neo S, McShaneand L, {et~al}. Breast cancer classification and
  prognosis based on gene expression profiles from a population-based study.
  {\it Proceedings of the National Academy of Sciences of the United States of
  America.} 2013\string; 100(18)\string: 10393-10398.

\bibitem{Burrell2013The}
Burrell R, McGranahan N, Bartekand J, {et~al}. The causes and consequences of
  genetic heterogeneity in cancer evolution. {\it Nature.} 2013\string;
  501(7467)\string: 338-345.

\bibitem{shen2009integrative}
Shen R, Olshen A, Ladanyi M. Integrative clustering of multiple genomic data
  types using a joint latent variable model with application to breast and lung
  cancer subtype analysis. {\it Bioinformatics.} 2009\string; 25(22)\string:
  2906--2912.

\bibitem{cancer2016comprehensive}
Linehan W, Spellman P, Ricketts C, {et~al}. Comprehensive molecular
  characterization of papillary renal-cell carcinoma. {\it New England Journal
  of Medicine.} 2016\string; 374(2)\string: 135--145.

\bibitem{chen2017multiplatform}
Chen F, Zhan Y, Parra E, {et~al}. Multiplatform-based molecular subtypes of
  non-small-cell lung cancer. {\it Oncogene.} 2017\string; 36(10)\string:
  1384--1393.

\bibitem{bradner2017transcriptional}
Bradner J, Hnisz D, Richard A. Transcriptional addiction in cancer. {\it Cell.}
  2017\string; 168(4)\string: 629--643.

\bibitem{kagohara2018epigenetic}
Kagohara L, Stein-O'Brien G, Kelley D, {et~al}. Epigenetic regulation of gene
  expression in cancer: techniques, resources and analysis. {\it Briefings in
  functional genomics.} 2018\string; 17(1)\string: 49--63.

\bibitem{Dutta2011A}
Dutta B, Pusztai L, Qi Y, {et~al}. A network-based, integrative study to
  identify core biological pathways that drive breast cancer clinical subtypes.
  {\it British Journal of Cancer.} 2011\string; 106(6)\string: 1107-1116.

\bibitem{SafonovImmune}
Safonov A, Jiang T, Bianchini G. Immune gene expression is associated with
  genomic aberrations in breast cancer. {\it Cancer Research.} 2017\string;
  77(12)\string: 3317-3324.

\bibitem{Hyman2002Impact}
Hyman E, Kauraniemi P, Hautaniemi S. Impact of DNA Amplification on Gene
  Expression Patterns in Breast Cancer. {\it Cancer Research.} 2002\string;
  62(21)\string: 6240-6245.

\bibitem{Pollack2002Microarray}
Pollack J, Sorlie T, Perou C. Microarray analysis reveals a major direct role
  of DNA copy number alteration in the transcriptional program of human breast
  tumors. {\it Proceedings of the National Academy of Sciences of the United
  States of America.} 2002\string; 99(20)\string: 12963-12968.

\bibitem{Platzer2002Silence}
Platzer P, Upender M, Wilson K. Silence of Chromosomal Amplifications in Colon
  Cancer. {\it Cancer Research.} 2002\string; 62(4)\string: 1134-1138.

\bibitem{Liu2010Identifying}
Liu H, Brannon A, Reddy A. Identifying mRNA targets of microRNA dysregulated in
  cancer: with application to clear cell Renal Cell Carcinoma. {\it BMC Systems
  Biology.} 2010\string; 4\string: 51.

\bibitem{Lichtblau2017Comparative}
Yvonne L, Haldemann K, Lenze B, {et~al}. Comparative assessment of
  differential network analysis methods. {\it Briefings in Bioinformatics.}
  2017\string; 18(5)\string: 837-850.

\bibitem{Wu2018}
Wu C, Zhang Q, Jiang Y, {et~al}. Robust network-based analysis of the
  associations between (epi)genetic measurements. {\it Journal of Multivariate
  Analysis.} 2018\string; 168\string: 119-130.
  
\bibitem{Demichelis2012}
Demichelis F, Setlur S, Banerjee S, {et~al}. Identification of functionally active, 
low frequency copy number variants at 15q21.3 and 12q21.31 associated with 
prostate cancer risk. {\it Proceedings of the National Academy of Sciences of the United
  States of America.} 2012\string; 109(17)\string: 6686-6691.
 
 \bibitem{Kumaran2018}
 Kumaran M, Krishnan P, Cass C E, {et~al}. Breast cancer associated germline structural variants 
 harboring small noncoding RNAs impact post-transcriptional gene regulation. {\it Scientific Reports.} 2018\string; 8\string: 7529.  
  

\bibitem{Hastie1993}
Hastie T, Tibshirani R. Varying-coefficient models. {\it Journal of the Royal
  Statistical Society, Series B.} 1993\string; 55(4)\string: 757-796.

\bibitem{Ma2016Estimating}
Ma S, Huang J. Estimating subgroup-specific treatment effects via concave
  fusion. arXiv:1607.03717v2;  2016.

\bibitem{Shen2015Inference}
Shen J, He X. Inference for Subgroup Analysis With a Structured Logistic-Normal
  Mixture Model. {\it Journal of the American Statistical Association.}
  2015\string; 110(509)\string: 303-312.

\bibitem{Yamada2017Localized}
Makoto Y, Koh T, Tomoharu I. Localized Lasso for High-Dimensional Regression.
  {\it {\rm In} Proceedings of the 20th International Conference on Artificial
  Intelligence and Statistics (AISTATS).} 2017\string; 54\string: 325-333.

\bibitem{Shi2015Deciphering}
Shi X, Zhao Q, Huang J. Deciphering the associations between gene expression
  and copy number alteration using a sparse double Laplacian shrinkage
  approach. {\it Bioinformatics.} 2015\string; 31(24)\string: 3977-3983.

\bibitem{Peng2010Regularized}
Peng J, Zhu J, Bergamaschi A. Regularized multivariate regression for
  identifying master predictors with application to integrative genomics study
  of breast cancer. {\it Annals of Applied Statistics.} 2010\string;
  4(1)\string: 53-77.

\bibitem{Kong2014Exclusive}
Kong D, Fujimaki R, Liu J. Exclusive feature learning on arbitrary structures
  via l\textsubscript{1,2} norm. {\it {\rm In} Advances in Neural Information
  Processing Systems.} 2014\string; 2\string: 1655-1663.

\bibitem{Ma2017A}
Ma S, Huang J. A concave pairwise fusion approach to subgroup analysis. {\it
  Journal of the American Statistical Association.} 2017\string;
  112(517)\string: 410-423.

\bibitem{rand1971}
Rand W. Objective Criteria for the Evaluation of Clustering Methods. {\it
  Journal of the American Statistical Association.} 1971\string;
  66(336)\string: 846--850.

\bibitem{Sun2018Identification}
Sun Y, Jiang Y, Li Y. Identification of cancer omics commonality and difference
  via community fusion. {\it Statistics in Medicine.} 2019\string;
  38(7)\string: 1200--1212.

\bibitem{Sun2019An}
Sun Y, Sun Z, Jiang Y, {et~al}. An integrative sparse boosting analysis of
  cancer genomic commonality and difference. {\it Statistical methods in
  medical research.} 2019\string; 29(5)\string: 1325-1337.

\bibitem{Dewey2011RSEM}
Li B, Dewey C. RSEM: accurate transcript quantification from RNA-Seq data with
  or without a reference genome. {\it Bmc Bioinformatics.} 2011\string;
  12(1)\string: 323-323.

\bibitem{Jiang2016Integrated}
Jiang Y, Shi X, Zhao Q. Integrated analysis of multidimensional omics data on
  cutaneous melanoma prognosis. {\it Genomics.} 2016\string; 107(6)\string:
  223-230.

\bibitem{Hidetoshi2006The}
Sumimoto H, Imabayashi F, Iwata T, {et~al}. The BRAF-MAPK signaling pathway is
  essential for cancer-immune evasion in human melanoma cells. {\it The Journal
  of Experimental Medicine.} 2006\string; 203(7)\string: 1651-1656.

\bibitem{kunz2019modelling}
Kunz M, Vera J. Modelling of Protein Kinase Signaling Pathways in Melanoma and
  Other Cancers. {\it Cancers.} 2019\string; 11(4)\string: 465.

\bibitem{Gregory2015A}
Gregory K, Carroll R, Baladandayuthapani V, {et~al}. A Two-Sample Test for
  Equality of Means in High Dimension. {\it Journal of the American Statistical
  Association.} 2015\string; 110(510)\string: 837-849.

\bibitem{Teran2017Assisted}
Hidalgo T, Wu M, Ma S. Assisted clustering of gene expression data using ANCut.
  {\it Bmc Genomics.} 2017\string; 18(1)\string: 623-634.

\end{thebibliography}

\clearpage
\appendix
\renewcommand\arraystretch{1.0}
\subsection*{A. Details of the computational algorithm}
\renewcommand\theequation{A\arabic{equation}}
\setcounter{equation}{0}

For a given $(\eta,\delta,\theta,\gamma)$, the component of the Lagrange function $\mathcal{L}$ that depends on $\beta$ is
\begin{equation*}
L=\frac{1}{2}\lVert Y-Z\beta\rVert_2^2+\theta^{\top}(\beta-\delta)+
\gamma^\top(\mathcal{A}\beta-\eta)+\frac{\rho}{2}\Big(\lVert \beta-\delta\rVert_2^2+\lVert \mathcal{A}\beta-\eta\rVert_2^2\Big),
\end{equation*}
where $Y=(y^{(1)},\ldots,y^{(n)})^\top$, $Z=\hbox{diag}(X^{(1)},\ldots,X^{(n)})$, and $\mathcal{A}=D\otimes\mathbb{I}_p$. Here $D=\{e_i-e_{i+1},i=1,\ldots,n-1\}^\top$ with $e_i$ being a length-$n$ column vector whose $i$th element equals 1 and the rest equal 0, $\mathbb{I}_p$ is a $p\times p$ identity matrix, and $\otimes$ denotes the Kronecker product. Setting $\frac{\partial L}{\partial \beta}=0$ leads to the update equation for $\beta$. For a given $(\eta(t),\delta(t),\theta(t),\gamma(t))$, where $\cdot(t)$ denotes the estimate at the $t$th update, the update for $\beta(t+1)$ is

\begin{equation}
\label{eq:beta}	
\beta(t+1)=\Big[Z^\top Z+\rho(\mathcal{A}^\top\mathcal{A}+\mathbb{I}_{np})\Big]^{-1}\Big[Z^\top Y-\mathcal{A}^\top\gamma(t)-\theta(t)+\rho(\mathcal{A}^\top\eta(t)+\delta(t))\Big].
\end{equation}
For $\eta^{(i)}$, the relevant component in the Lagrange function is
\begin{equation*}
\lambda_2 \lVert \eta^{(i)}\rVert_2+\frac{\rho}{2}\lVert \beta^{(i)}-\beta^{(i+1)}-\eta^{(i)}+\frac{\gamma^{(i)}}{\rho}\rVert_2^2,
\end{equation*}
which leads to the closed-form solution of $\eta^{(i)}$, denoted as $\hat\eta^{(i)}$:
\begin{equation*}
\hat\eta^{(i)}=\Big(1-\frac{\lambda_2}{\rho \lVert \beta^{(i)}-\beta^{(i+1)}+\frac{\gamma^{(i)}}{\rho}\rVert}_2\Big)_+
\Big(\beta^{(i)}-\beta^{(i+1)}+\frac{\gamma^{(i)}}{\rho}\Big).
\end{equation*}
Then we obtain the update of $\eta^{(i)}$ at the $t$th iteration:
\begin{equation}
	\label{eq:eta}
\eta^{(i)}(t)  = \Big(1-\frac{\lambda_2}{\rho\lVert \beta^{(i)}(t+1)-\beta^{(i+1)}(t+1)+\frac{\gamma^{(i)}(t)}
{\rho}\rVert_2}\Big)_+ \Big(\beta^{(i)}(t+1)-\beta^{(i+1)}(t+1)+\frac{\gamma^{(i)}(t)}{\rho}\Big).
\end{equation}
The component of the Lagrange function that depends on $\delta$ is
\begin{equation}
\label{eq:delta1}
\lambda_1\sum\limits_{i=1}^n\lVert\delta^{(i)}\rVert_1^2+
\frac{\rho}{2}\lVert\beta-\delta+\frac{\theta}{\rho}\rVert_2^2.
\end{equation}
Minimizing (\ref{eq:delta1}) leads to the update
\begin{equation}
\label{eq:delta}
\delta^{(t+1)} = (\mathbb{I}_{np}+\frac{2\lambda_1}{\rho}\mathcal{H})^{-1}
(\beta^{t+1}+\frac{\theta^t}{\rho}).
\end{equation}
Here $\mathcal{H}$ is a $np\times np$ diagonal matrix with the $s$th diagonal element as $\frac{\lVert\delta^{(\lceil \frac{s}{p} \rceil)}\rVert_1}{|[\delta]_s|}$, where $\lceil x \rceil$ represents the smallest integer which is greater than or equal to $x$, and $[\delta]_s$ denotes the $s$th element of vector $\delta$. Finally, the updates of $\gamma$ and $\theta$ are given by
\begin{equation}
\label{eq:theta}
\theta^{t+1}=\theta^t+\rho(\beta^{t+1}-\delta^{t+1}),
\end{equation}
and
\begin{equation}
\label{eq:gamma}
\gamma^{t+1}=\gamma^t+\rho(\mathcal{A}\beta^{t+1}-\eta^{t+1}). 
\end{equation}
The overall algorithm is summarized in Algorithm \ref{algo}. Figure \ref{fig:time} shows the primal residual by iteration for a random replicate under setting S1, balanced structure, and $(\rho,\sigma^2)=(0.3,1)$. The convergence criterion is satisfied after 49 iterations. We have also examined a few other replicates and observed similar patterns. 

\noindent{\bf Tuning parameter selection} The proposed MSF approach involves two tuning parameters $\lambda_1$ and $\lambda_2$. $\lambda_1$ controls the sparsity of variable selection, and $\lambda_2$ controls the structure of subgroups. We apply a grid search for $\lambda_1$ and $\lambda_2$ with $\lambda_1\in\{0,0.1,\ldots,1\}$ and $\lambda_2\in\{10,50,100,150,200,250\}$ in our numerical study and select the optimal tuning parameters by minimizing BIC:  
\begin{equation*}
\text{BIC}=\sum\limits_{j=1}^q\log\Big[\sum\limits_{i=1}^n\Big(y_j^{(i)}-X^{(i)}\hat\beta_j^{(i)}\Big)^2\Big]+\frac{\log n}{n}df,
\end{equation*}
where $df$ is the total number of nonzero regression coefficients in all distinct values of $\{\hat{\bm B}^{(1)},\ldots, \hat{\bm B}^{(n)}\}$.  

\begin{algorithm}[h]
\caption{Algorithm for solving (3)}
\label{algo}
\begin{algorithmic}[1]
\State Set $t=0$. For $j=1,\ldots,q$, initialize $\beta_j(0)=0$, and $\eta_j(0)=\mathcal{A}\beta_j(0)$, $\delta_j(0)=\beta_j(0)$, $\theta_j(0)=0$, and $\gamma_j(0)=0$;
\Repeat
   \For {$j=1,\ldots,q$}
      \State Update $\beta_j(t+1)$ via (\ref{eq:beta});
      \State Update  $\eta_j(t+1)$ via (\ref{eq:eta});
      \State Update $\delta_j(t+1)$ via (\ref{eq:delta});
      \State Update $\theta_j(t+1)$ and $\gamma_j(t+1)$ via  (\ref{eq:theta}) and (\ref{eq:gamma}), respectively.
     \EndFor
  \State $t=t+1$;
  \Until the primal residual $\max\limits_{j\in\{1,\ldots,q\}}\{\parallel \mathcal{A}\beta_j(t+1)-\delta_j(t+1)\parallel_2,\parallel     \beta_j(t+1)-\theta_j(t+1)\parallel_2\}$ is less than a predefined threshold, e.g. $10^{-6}$;
\State \Return the estimates of $\beta_j$ ($j=1,\ldots,q$) at convergence.
\end{algorithmic}
\end{algorithm}

\clearpage

\subsection*{B. Alternative approaches used in numerical study}
Three alternative approaches are adopted in our numerical study: Response-based clustering (RespClust), Residual-based clustering (ResiClust), and Ridge fusion clustering (RiFuClust). All approaches follow a similar strategy: first samples are grouped, and then Lasso is applied to each subgroup. To make the comparison with the proposed approach fair, we also incorporate information of the guiding biomarker in grouping. Specifically, as with the proposed MSF approach, first, the samples are sorted according to the guiding biomarker. Second, the difference in a certain measure between any two adjacent samples, which is broadly referred to as distance, is computed. Specifically, this measure is chosen as response variable with RespClust, residual with ResiClust, and estimated coefficient matrix with RiFuClust. Third, all distances beyond a given threshold are identified as cutoffs. In our numerical study, the threshold is set as 0.1 (which overall generates better results than other values). Finally, to avoid generating subgroups with sizes too small, we remove some cutoffs so that there are at least 20\% samples between two adjacent cutoffs.

\newpage
\subsection*{C. Additional numerical results}
\renewcommand\thetable{C\arabic{table}}
\setcounter{table}{0}
\renewcommand\arraystretch{1.0}
\begin{table}[!htbp]
 \caption{Simulation under setting S2. In each cell, mean(sd).}
\label{sim:S2}
 \begin{center}
    \setlength{\tabcolsep}{5pt}
    \renewcommand{\arraystretch}{1.1}
	\begin{tabular}{ccccccccccc}
\hline
    &  &  & & \multicolumn{2}{c}{Heterogeneity identification} & & \multicolumn{2}{c}{Variable selection} \\
	\cline{5-6} \cline{8-9}
   Structure & \multicolumn{2}{c}{($\rho$,$\sigma^2$)} & Method &  Num &  Rand	   && TPR	    &FPR 	& EMSE & PMSE  \\\hline
\multirow{16}{*}{Balanced} &
\multicolumn{2}{c}{\multirow{4}{*}{(0.3,1)}}
&	    MSF         &	3.10(0.30)  &	0.958(0.039)    &&	89.3\%(12.4\%)	&	1.7\%(0.5\%)	&	0.017(0.008)	&	3.952(1.883)	\\
& & &	RespClust	&	3.36(0.63)  &	0.836(0.061)  	&&	76.6\%(10.8\%)	&	1.8\%(0.5\%)	&	0.046(0.018)	&	10.664(4.074)	\\
& & &	ResiClust	&	3.34(0.63) 	&	0.843(0.061) 	&&	77\%(10.9\%)	&	1.8\%(0.6\%)	&	0.044(0.019)	&	10.248(4.212)	\\
& & &	RiFuClust 	&	3.62(0.49) 	&	0.907(0.044) 	&&	92\%(8.3\%)	&	2.1\%(0.6\%)	&	0.023(0.009)	&	4.628(1.862) 	\\
\cline{2-11}
& \multicolumn{2}{c}{\multirow{4}{*}{(0.3,3)}}
&	    MSF	        &	3.06(0.24) 	&	0.963(0.032) 	&&	87.1\%(12\%)	&	1.8\%(0.4\%)	&	0.021(0.007)	&	6.272(1.594) 	\\
& & &	RespClust	&	3.28(0.64)  &	0.845(0.063) 	&&	75.5\%(10.9\%)	&	1.9\%(0.5\%)	&	0.048(0.017)	&	12.46(3.79)	\\
& & &	ResiClust	&	3.30(0.65)  &	0.841(0.061) 	&&	75.3\%(10.1\%)	&	1.9\%(0.5\%)	&	0.049(0.018)	&	12.607(3.925) 	\\
& & &   RiFuClust 	&	3.66(0.48) 	&	0.903(0.045) 	&&	88.2\%(9.4\%)	&	2.2\%(0.5\%)	&	0.032(0.011)	&	7.814(2.155)		\\
\cline{2-11}
& \multicolumn{2}{c}{\multirow{4}{*}{(0.8,1)}}
&	    MSF	        &	3.22(0.42) 	&	0.966(0.025) 	&&	92.4\%(6.4\%)	&	0.5\%(0.1\%)	&	0.016(0.01)	&	6.388(6.322)	\\
& & &	RespClust	&	4.42(0.61) 	&	0.843(0.035) 	&&	75.3\%(6.5\%)	&	0.8\%(0.3\%)	&	0.034(0.009)	&	15.132(4.782)	\\
& & &	ResiClust	&	4.38(0.57) 	&	0.846(0.037) 	&&	75.5\%(6.6\%)	&	0.7\%(0.3\%)	&	0.035(0.009)	&	15.149(4.901)	\\
& & &	RiFuClust 	&	5.72(0.45) 	&	0.837(0.019)	&&	94.6\%(5.1\%)	&	0.9\%(0.2\%)	&	0.018(0.007)	&	5.73(2.363)	\\
\cline{2-11}
& \multicolumn{2}{c}{\multirow{4}{*}{(0.8,3)}}
&	    MSF         &	3.22(0.42) 	&	0.964(0.027) 	        &&	89.9\%(6.9\%)	&	0.6\%(0.1\%)	&	0.025(0.009)	&	8.348(4.096) 	\\
& & &	RespClust	&	4.40(0.57) 	&	0.848(0.034) 	&&	74.1\%(6.2\%)	&	0.7\%(0.2\%)	&	0.043(0.009)	&	17.518(5.037) 	\\
& & &	ResiClust	&	4.38(0.57) 	&	0.85(0.035) 	&&	74.5\%(5.9\%)	&	0.7\%(0.2\%)	&	0.044(0.009)	&	18.123(4.986)	\\
& & &	RiFuClust 	&	5.58(0.57) 	&	0.842(0.025) 	&&	90.2\%(5.9\%)	&	1\%(0.2\%)	&	0.032(0.009)	&	7.844(2.729)	\\\hline
\multirow{16}{*}{Unbalanced} &
\multicolumn{2}{c}{\multirow{4}{*}{(0.3,1)}}
&	    MSF	        &	3.10(0.30) 	&	0.947(0.041) 	&&	92.9\%(9.7\%)	&	1.7\%(0.6\%)	&	0.017(0.008)	&	3.674(1.463) 	\\
& & &	RespClust	&	3.06(0.77) 	&	0.763(0.084) 	&&	56.3\%(19\%)	&	1.6\%(0.7\%)	&	0.058(0.023)	&	14.054(5.507) 	\\
& & &	ResiClust	&	2.96(0.70) 	&	0.76(0.091) 	&&	55.2\%(19.1\%)	&	1.5\%(0.7\%)	&	0.059(0.023)	&	14.373(5.489)	\\
& & &	RiFuClust 	&	2.46(0.50) 	&	0.79(0.098) 	&&	90.6\%(11.7\%)	&	2.2\%(1.1\%)	&	0.043(0.017)	&	8.87(3.547)	\\
\cline{2-11}
& \multicolumn{2}{c}{\multirow{4}{*}{(0.3,3)}}
&	    MSF	        &	3.38(0.57) 	&	0.92(0.046) 	&&	88\%(12.3\%)	&	2\%(0.7\%)	&	0.025(0.01)	&	6.479(2.466)	\\
& & &	RespClust	&	4.62(0.57) 	&	0.82(0.035)    &&	74\%(9.1\%)	&	2.3\%(0.5\%)	&	0.046(0.014)	&	10.551(2.724) 	\\
& & &   ResiClust	&	4.68(0.59) 	&	0.822(0.035) 	&&	72.8\%(9.7\%)	&	2.3\%(0.5\%)	&	0.045(0.014)	&	10.524(2.89)	\\
& & &	RiFuClust 	&	4.24(0.69) 	&	0.878(0.043) 	&&	86.6\%(6.3\%)	&	2.3\%(0.5\%)	&	0.034(0.013)	&	7.685(2.337) 	\\
\cline{2-11}
& \multicolumn{2}{c}{\multirow{4}{*}{(0.8,1)}}
&	    MSF	        &	3.36(0.48) 	&	0.951(0.04) 	&&	90.4\%(8.4\%)	&	0.6\%(0.2\%)	&	0.016(0.007)	&	5.018(2.479) 	\\
& & &   RespClust   &	4.90(0.65) 	&	0.816(0.033) 	&&	78.4\%(6.8\%)	&	0.8\%(0.2\%)	&	0.036(0.013)	&	14.765(6.469)	\\
& & &	ResiClust	&	4.88(0.52) 	&	0.817(0.034) 	&&	77.3\%(7.9\%)	&	0.8\%(0.3\%)	&	0.035(0.012)	&	14.717(6.085) 	\\
& & &	RiFuClust 	&	5.06(0.31) 	&	0.847(0.015) 	&&	93.5\%(5.6\%)	&	0.9\%(0.2\%)	&	0.019(0.008)	&	5.17(2.614)	\\
\cline{2-11}
& \multicolumn{2}{c}{\multirow{4}{*}{(0.8,3)}}
&	    MSF	        &	3.48(0.65) 	&	0.941(0.045) 	&&	88.2\%(8.6\%)	&	0.6\%(0.2\%)	&	0.022(0.006)	&	7.52(2.633)	\\
& & &   RespClust	&	5.88(0.56) 	&	0.794(0.018) 	&&	78.2\%(5.3\%)	&	0.9\%(0.2\%)	&	0.042(0.012)	&	16.586(6.592) 	\\
& & &   ResiClust	&	5.86(0.57) 	&	0.795(0.02) 	&&	78.3\%(4.6\%)	&	0.9\%(0.2\%)	&	0.039(0.01)	&	15.353(4.946) 	\\
& & &	RiFuClust 	&	6.68(0.59) 	&	0.8(0.018) 	&&	89.8\%(4.5\%)	&	1.1\%(0.2\%)	&	0.026(0.005)	&	7.053(2.248) 	\\\hline
 	\end{tabular}
\end{center}
\end{table}

\clearpage
\renewcommand\arraystretch{1.0}
\begin{table}
 \caption{Simulation under setting S3. In each cell, mean(sd).}
\label{sim:S3}
 \begin{center}
    \setlength{\tabcolsep}{5pt}
    \renewcommand{\arraystretch}{1.0}
	\begin{tabular}{ccccccccccc}
\hline
    &  &  & & \multicolumn{2}{c}{Heterogeneity identification} & & \multicolumn{2}{c}{Variable selection} \\
	\cline{5-6} \cline{8-9}
   Structure & \multicolumn{2}{c}{($\rho$,$\phi$)} & Method & Num	& Rand	   && TPR	    &FPR 	& EMSE & PMSE  \\\hline
\multirow{16}{*}{Balanced} &
\multicolumn{2}{c}{\multirow{4}{*}{(0.3,0.3)}}
&	    MSF         &	3.00(0.00) 	&	0.984(0.009)    &&	98.8\%(0.7\%)	&	0.6\%(0.1\%)	&	0(0)	&	0.326(0.032)	\\
& & &	RespClust	&	3.56(0.58) 	&	0.859(0.046)  	&&	86.3\%(6.1\%)	&	0.7\%(0.1\%)	&	0.003(0.001)	&	0.721(0.161)	\\
& & &	ResiClust	&	3.50(0.65) 	&	0.857(0.051) 	&&	86.2\%(6.3\%)	&	0.7\%(0.1\%)	&	0.003(0.001)	&	0.734(0.189)	\\
& & &	RiFuClust 	&	3.30(0.51) 	&	0.956(0.044) 	&&	96\%(4.6\%)	    &	0.6\%(0.1\%)	&	0.001(0.001)	&	0.391(0.104) 	\\
\cline{2-11}
& \multicolumn{2}{c}{\multirow{4}{*}{(0.3,0.8)}}
&	    MSF	        &	3.04(0.20) 	&	0.98(0.015) 	&&	98.6\%(0.9\%)	&	0.5\%(0.1\%)	&	0.001(0)	    &	0.33(0.036) 	\\
& & &	RespClust	&	4.50(0.58) 	&	0.841(0.043) 	&&	84.8\%(6.6\%)	&	0.8\%(0.1\%)	&	0.002(0.001)	&	0.682(0.145)	\\
& & &	ResiClust	&	4.44(0.58) 	&	0.846(0.043) 	&&	85.3\%(6.7\%)	&	0.8\%(0.1\%)	&	0.002(0.001)	&	0.668(0.144) 	\\
& & &   RiFuClust 	&	5.86(0.35) 	&	0.83(0.011) 	&&	96\%(3.7\%)	    &	1\%(0.1\%)	    &	0.001(0)	    &	0.385(0.078)	\\
\cline{2-11}
& \multicolumn{2}{c}{\multirow{4}{*}{(0.8,0.3)}}
&	    MSF	        &	3.06(0.24) 	&	0.985(0.013) 	&&	99\%(0.4\%)	&	0.3\%(0.1\%)	&	0.001(0)	&	0.358(0.039)	\\
& & &	RespClust	&	4.44(0.64) 	&	0.839(0.045) 	&&	84.1\%(5.3\%)	&	0.5\%(0.1\%)	&	0.003(0.001)	&	1.046(0.225)	\\
& & &	ResiClust	&	4.48(0.58) 	&	0.839(0.042) 	&&	84.1\%(5.5\%)	&	0.6\%(0.1\%)	&	0.003(0.001)	&	1.042(0.235)	\\
& & &	RiFuClust 	&	6.00(0.00) 	&	0.83(0.001) 	&&	98.6\%(0.5\%)	&	0.5\%(0\%)	    &	0.001(0)	&	0.351(0.026)	\\
\cline{2-11}
& \multicolumn{2}{c}{\multirow{4}{*}{(0.8,0.8)}}
&	    MSF         &	3.02(0.14) 	&	0.988(0.009) 	&&	99.1\%(0.4\%)	&	0.3\%(0.1\%)	&	0(0)	&	0.351(0.037) 	\\
& & &	RespClust	&	4.42(0.61) &	0.838(0.044) 	&&	84.2\%(5.8\%)	&	0.5\%(0.1\%)	&	0.003(0.001)	&	1.059(0.272) 	\\
& & &	ResiClust	&	4.40(0.61) 	&	0.846(0.044) 	&&	85\%(6.2\%)	&	0.5\%(0.1\%)	&	0.003(0.001)	&	1.015(0.281)	\\
& & &	RiFuClust 	&	6.00(0.00) 	&	0.83(0.001) 	&&	98.6\%(0.5\%)	&	0.5\%(0.1\%)	&	0.001(0)	&	0.348(0.026)	\\\hline
\multirow{16}{*}{Unbalanced} &
\multicolumn{2}{c}{\multirow{4}{*}{(0.3,0.3)}}
&	    MSF	        &	3.02(0.14) 	&	0.983(0.015) 	&&	98.7\%(0.8\%)	&	0.5\%(0.1\%)	&	0.001(0)	&	0.328(0.034) 	\\
& & &	RespClust	&	4.34(0.63) 	&	0.822(0.036) 	&&	84.8\%(5.4\%)	&	0.8\%(0.1\%)	&	0.003(0.001)	&	0.704(0.152) 	\\
& & &	ResiClust	&	4.42(0.61) 	&	0.822(0.035) 	&&	85.6\%(5.6\%)	&	0.8\%(0.1\%)	&	0.002(0.001)	&	0.685(0.153)	\\
& & &	RiFuClust 	&	5.00(0.00) 	&	0.848(0.008) 	&&	97\%(19\%)	&	0.8\%(0.1\%)	&	0.001(0)	&	0.389(0.052)	\\
\cline{2-11}
& \multicolumn{2}{c}{\multirow{4}{*}{(0.3,0.8)}}
&	    MSF	        &	3.04(0.20) 	&	0.98(0.019) 	&&	98.7\%(0.8\%)	&	0.5\%(0.1\%)	&	0.001(0)	&	0.328(0.035)	\\
& & &	RespClust	&	4.42(0.57)	&	0.816(0.032)    &&	     84.6\%(6.1\%)	&	0.8\%(0.1\%)	&	0.003(0.001)	&	0.717(0.146) 	\\
& & &   ResiClust	&	4.46(0.61) 	&	0.817(0.033) 	&&	84.9\%(6\%)	&	0.8\%(0.1\%)	&	0.003(0.001)	&	0.707(0.135)	\\
& & &	RiFuClust 	&	5.00(0.00) 	&	0.852(0.007) 	&&	97\%(2.1\%)	&	0.9\%(0.1\%)	&	0.001(0)	&	0.372(0.051) 	\\
\cline{2-11}
& \multicolumn{2}{c}{\multirow{4}{*}{(0.8,0.3)}}
&	    MSF	        &	3.12(0.33) 	&	0.979(0.025)	&&	99\%(0.5\%)	&	0.3\%(0.1\%)	&	0.001(0)	&	0.363(0.031) 	\\
& & &   RespClust   &	4.44(0.67) 	&	0.815(0.032) 	&&	84.6\%(5.6\%)	&	0.5\%(0.1\%)	&	0.003(0.001)	&	1.092(0.316)	\\
& & &	ResiClust	&	4.38(0.64) 	&	0.812(0.03) 	&&	83.6\%(5\%)	&	0.5\%(0.1\%)	&	0.003(0.001)	&	1.135(0.289) 	\\
& & &	RiFuClust 	&	5.00(0.00) 	&	0.86(0.001) 	&&	98.5\%(0.5\%)	&	0.5\%(0\%)	&	0.001(0)	&	0.355(0.028)	\\
\cline{2-11}
& \multicolumn{2}{c}{\multirow{4}{*}{(0.8,0.8)}}
&	    MSF	        &	3.00(0.00) 	&	0.989(0.003)	&&	99.1\%(0.3\%)	&	0.3\%(0.1\%)	&	0.001(0)	&	0.356(0.029)	\\
& & &   RespClust	&	4.46(0.68) 	&	0.818(0.036) 	&&	84.9\%(5.7\%)	&	0.5\%(0.1\%)	&	0.003(0.001)	&	1.067(0.299) 	\\
& & &   ResiClust	&	4.44(0.64) 	&	0.815(0.03) 	&&	84.2\%(5.5\%)	&	0.5\%(0.1\%)	&	0.003(0.001)	&	1.098(0.306) 	\\
& & &	RiFuClust 	&	5.00(0.00) 	&	0.86(0.001) 	&&	98.5\%(0.6\%)	&	0.5\%(0.1\%)	&	0.001(0)	&	0.357(0.027) 	\\\hline
 	\end{tabular}
\end{center}
\end{table}

\clearpage
\renewcommand\arraystretch{1.0}
\begin{table}
 \caption{Simulation under setting S4. In each cell, mean(sd).}
\label{sim:S4}
 \begin{center}
    \setlength{\tabcolsep}{5pt}
    \renewcommand{\arraystretch}{1.0}
	\begin{tabular}{ccccccccccc}
\hline
    &  &  & & \multicolumn{2}{c}{Heterogeneity identification} & & \multicolumn{2}{c}{Variable selection} \\
	\cline{5-6} \cline{8-9}
   Structure & \multicolumn{2}{c}{($\rho$,$\phi$)} & Method &  Num	& Rand	   && TPR	    &FPR 	& EMSE & PMSE  \\\hline
\multirow{16}{*}{Balanced} &
\multicolumn{2}{c}{\multirow{4}{*}{(0.3,0.3)}}
&	    MSF         &	3.00(0.00) 	&	0.989(0.002)    &&	98.5\%(0.9\%)	&	0.9\%(0.1\%)	&	0.004(0.001)	&	0.867(0.133)	\\
& & &	RespClust	&	4.48(0.54) 	&	0.846(0.027)  	&&	81.2\%(6.5\%)	&	1.5\%(0.3\%)	&	0.018(0.005)	&	3.667(0.936)	\\
& & &	ResiClust	&	4.46(0.54) 	&	0.848(0.031) 	&&	82.1\%(6.7\%)	&	1.5\%(0.3\%)	&	0.018(0.005)	&	3.588(0.952)	\\
& & &	RiFuClust 	&	3.02(0.14) 	&	0.819(0.013) 	&&	95\%(1\%)	    &	1.4\%(0.1\%)	&	0.017(0.001)	&	3.288(0.28) 	\\
\cline{2-11}
& \multicolumn{2}{c}{\multirow{4}{*}{(0.3,0.8)}}
&	    MSF	        &	3.06(0.31) 	&	0.986(0.017) 	&&	98.5\%(0.9\%)	&	1\%(0.2\%)	&	0.004(0.001)	&	0.875(0.162) 	\\
& & &	RespClust	&	4.48(0.58) 	&	0.85(0.03) 	&&	83.1\%(5.6\%)	&	1.5\%(0.3\%)	&	0.017(0.004)	&	3.366(0.819)	\\
& & &	ResiClust	&	4.46(0.58) 	&	0.848(0.033) 	&&	82.6\%(5.7\%)	&	1.5\%(0.3\%)	&	0.017(0.005)	&	3.462(0.933) 	\\
& & &   RiFuClust 	&	4.50(0.51) 	&	0.895(0.029) 	&&	96.4\%(1.4\%)	&	1.5\%(0.2\%)	&	0.006(0.002)	&	1.203(0.342)	\\
\cline{2-11}
& \multicolumn{2}{c}{\multirow{4}{*}{(0.8,0.3)}}
&	    MSF	        &	3.04(0.20) 	&	0.988(0.011) 	&&	97.2\%(1.2\%)	&	0.5\%(0.1\%)	&	0.004(0.001)	&	1.38(0.291)	\\
& & &	RespClust	&	4.28(0.45) 	&	0.856(0.033) 	&&	79.5\%(5.9\%)	&	0.7\%(0.1\%)	&	0.019(0.005)	&	6.861(1.924)	\\
& & &	ResiClust	&	4.24(0.43) 	&	0.856(0.033) 	&&	79\%(6.4\%)	&	0.7\%(0.1\%)	&	0.019(0.005)	&	6.942(2.001)	\\
& & &	RiFuClust 	&	5.10(0.30) 	&	0.877(0.016) 	&&	95.9\%(0.9\%)	&	0.8\%(0.1\%)	&	0.005(0.001)	&	1.355(0.207)	\\
\cline{2-11}
& \multicolumn{2}{c}{\multirow{4}{*}{(0.8,0.8)}}
&	    MSF         &	3.06(0.24) 	&	0.986(0.013) 	&&	96.6\%(1.2\%)	&	0.5\%(0.1\%)	&	0.005(0.001)	&	1.512(0.288) 	\\
& & &	RespClust	&	4.38(0.49) 	&	0.848(0.029) 	&&	79.8\%(5.6\%)	&	0.7\%(0.1\%)	&	0.019(0.005)	&	6.791(1.963) 	\\
& & &	ResiClust	&	4.40(0.49) 	&	0.849(0.03) 	&&	80.5\%(5.1\%)	&	0.8\%(0.1\%)	&	0.019(0.005)	&	6.67(1.972)	\\
& & &	RiFuClust 	&	5.14(0.35) 	&	0.875(0.019) 	&&	95.6\%(0.9\%)	&	0.8\%(0.1\%)	&	0.005(0.001)	&	1.505(0.226)	\\\hline
\multirow{16}{*}{Unbalanced} &
\multicolumn{2}{c}{\multirow{4}{*}{(0.3,0.3)}}
&	    MSF	        &	3.04(0.20) 	&	0.986(0.016) 	&&	98\%(1.4\%)	&	1\%(0.2\%)	&	0.004(0.001)	&	0.935(0.182) 	\\
& & &	RespClust	&	4.54(0.50) 	&	0.825(0.03) 	&&	82\%(7.6\%)	&	1.5\%(0.2\%)	&	0.019(0.006)	&	3.8(1.351) 	\\
& & &	ResiClust	&	4.52(0.54) 	&	0.825(0.034) 	&&	82.6\%(7.8\%)	&	1.5\%(0.2\%)	&	0.019(0.008)	&	3.891(1.544)	\\
& & &	RiFuClust 	&	4.00(0.00) 	&	0.925(0.014) 	&&	96.2\%(1.4\%)	&	1.3\%(0.1\%)	&	0.005(0.002)	&	1.091(0.336)	\\
\cline{2-11}
& \multicolumn{2}{c}{\multirow{4}{*}{(0.3,0.8)}}
&	    MSF	        &	3.08(0.27) 	&	0.985(0.02) 	&&	98.4\%(0.9\%)	&	1\%(0.2\%)	&	0.004(0.001)	&	0.868(0.148)	\\
& & &	RespClust	&	5.76(0.69) 	&	0.806(0.028)    &&	84.8\%(5.6\%)	&	1.8\%(0.2\%)	&	0.015(0.005)	&	2.92(1.008) 	\\
& & &   ResiClust	&	5.66(0.63) 	&	0.806(0.029) 	&&	83.6\%(5.2\%)	&	1.7\%(0.2\%)	&	0.016(0.005)	&	3.121(1.051)	\\
& & &	RiFuClust 	&	4.46(0.50) 	&	0.912(0.027) 	&&	93.8\%(2.9\%)	&	1.4\%(0.2\%)	&	0.007(0.003)	&	1.363(0.497) 	\\
\cline{2-11}
& \multicolumn{2}{c}{\multirow{4}{*}{(0.8,0.3)}}
&	    MSF	        &	3.22(0.42) 	&	0.974(0.03) 	&&	97.1\%(1.2\%)	&	0.5\%(0.1\%)	&	0.005(0.001)	&	1.388(0.252) 	\\
& & &   RespClust   &	5.80(0.53) 	&	0.799(0.021) 	&&	83.6\%(4.9\%)	&	0.9\%(0.1\%)	&	0.016(0.004)	&	5.232(1.645)	\\
& & &	ResiClust	&	5.88(0.48) 	&	0.797(0.021) 	&&	83.6\%(5.1\%)	&	0.9\%(0.1\%)	&	0.016(0.004)	&	5.268(1.572) 	\\
& & &	RiFuClust 	&	6.00(0.00) 	&	0.843(0)    	&&	95.7\%(0.8\%)	&	0.9\%(0.1\%)	&	0.005(0)	&	1.381(0.176)	\\
\cline{2-11}
& \multicolumn{2}{c}{\multirow{4}{*}{(0.8,0.8)}}
&	    MSF	        &	3.22(0.42) 	&	0.974(0.031)	&&	97\%(1.5\%)	&	0.6\%(0.1\%)	&	0.005(0.001)	&	1.404(0.297)	\\
& & &   RespClust	&	5.76(0.52) 	&	0.798(0.021) 	&&	83.9\%(4.8\%)	&	0.9\%(0.1\%)	&	0.017(0.005)	&	5.558(1.692) 	\\
& & &   ResiClust	&	5.82(0.48) 	&	0.799(0.022) 	&&	84.5\%(4.9\%)	&	0.9\%(0.1\%)	&	0.016(0.004)	&	5.307(1.626) 	\\
& & &	RiFuClust 	&	6.00(0.00) 	&	0.843(0) 	&&	95.6\%(1\%)	&	0.9\%(0.1\%)	&	0.006(0.001)	&	1.424(0.228) 	\\\hline
 	\end{tabular}
\end{center}
\end{table}

\clearpage
\renewcommand\arraystretch{1.0}
\begin{table}[]
\centering
\caption{Simulation under setting S4 with nonzero coefficients generated from truncated normal distributions. In each cell, mean(sd).}
\label{sim:S4_2}
    \setlength{\tabcolsep}{5pt}
    \renewcommand{\arraystretch}{1.0}
	\begin{tabular}{ccccccccccc}
\hline
    &  &  & & \multicolumn{2}{c}{Heterogeneity identification} & & \multicolumn{2}{c}{Variable selection} \\
	\cline{5-6} \cline{8-9}
   Structure & \multicolumn{2}{c}{($\rho$,$\phi$)} & Method &  Num	& Rand	   && TPR	    &FPR 	& EMSE & PMSE  \\\hline
\multirow{4}{*}{Balanced} &
\multicolumn{2}{c}{\multirow{4}{*}{(0.3,0.3)}}
& MSF                  & 3.02(0.21)                         & 0.977(0.014)              && 97.8\%(1.1\%)            & 0.8\%(0.1\%)            & 0.005(0.001)         & 0.871(0.122)         \\
                            &&                                      & RespClust            & 4.44(0.56)              & 0.849(0.031)              && 81.1\%(7.1\%)            & 1.4\%(0.3\%)            & 0.019(0.007)         & 3.742(1.029)         \\
                            &&                                      & ResiClust            & 4.42(0.53)              & 0.851(0.033)              && 81.7\%(7.3\%)            & 1.5\%(0.2\%)            & 0.018(0.006)         & 3.634(0.984)         \\
                            &&                                      & RiFuClust            & 3.07(0.18)              & 0.868(0.016)              && 95.3\%(1.2\%)            & 1.6\%(0.3\%)            & 0.008(0.005)         & 3.211(0.291)         \\ \midrule
\multirow{4}{*}{Unbalanced} & 
\multicolumn{2}{c}{\multirow{4}{*}{(0.3,0.3)}}
& MSF                  & 3.05(0.22)              & 0.980(0.018)              && 97.9\%(1.6\%)            & 1.5\%(0.3\%)            & 0.005(0.002)         & 0.948(0.199)         \\
                            &&                                      & RespClust            & 4.42(0.56)              & 0.827(0.033)              && 82.3\%(6.7\%)            & 1.7\%(0.2\%)            & 0.021(0.005)         & 3.871(1.274)         \\
                            &&                                      & ResiClust            & 4.44(0.54)              & 0.826(0.035)              && 82.8\%(7.3\%)            & 1.6\%(0.3\%)            & 0.020(0.004)         & 3.741(1.309)         \\
                            &&                                      & RiFuClust            & 4.05(0.43)              & 0.921(0.014)              && 96.4\%(1.9\%)            & 2.8\%(0.3\%)           & 0.005(0.003)         & 1.066(0.340)         \\ \bottomrule
\end{tabular}
\end{table}

\clearpage
\renewcommand\arraystretch{1.0}
 \begin{table}
\caption{Data analysis: comparison of subgrouping structures. In each cell, $p$-value.}
 \begin{center}
    \setlength{\tabcolsep}{2.5pt}
    \renewcommand{\arraystretch}{1.0}
	\begin{tabular}{lccc}
    \hline
 & RespClust & ResiClust & RiFuClust \\ \hline
 \multicolumn{4}{c}{Melanoma data}     \\
MSF  & $<$2.2e-16 & $<$2.2e-16 & $<$2.2e-16 \\
RespClust &  & 0.069 & $<$2.2e-16 \\
ResiClust &  &  &  $<$2.2e-16 \\ \hline
 \multicolumn{4}{c}{Stomach cancer data}\\
MSF &$<$2.2e-16   & $<$2.2e-16 & $<$2.2e-16  \\
RespClust &   & 0.049 & $<$2.2e-16 \\
ResiClust &  & & $<$2.2e-16  \\
\hline
 	\end{tabular}
\end{center}
\label{tab:pval}
\end{table}

\renewcommand\arraystretch{1.0}
 \begin{table}
\caption{Data analysis: comparison of variable selection results. In each cell, number of overlapping identification. }
 \begin{center}
    \setlength{\tabcolsep}{2.5pt}
    \renewcommand{\arraystretch}{1.0}
	\begin{tabular}{lcccc}
    \hline
 & MSF & RespClust & ResiClust & RiFuClust \\ \hline
 \multicolumn{5}{c}{Melanoma data}     \\
MSF & 4309 & 1464 & 1620 & 1482 \\
RespClust &  & 5281 & 3038 & 1863 \\
ResiClust &  &  & 6245 & 2354 \\
RiFuClust &  &  &  & 5442 \\ \hline
 \multicolumn{5}{c}{Stomach cancer data}\\
MSF & 11254 & 6788 & 7388 & 4668 \\
RespClust &  & 11802 & 8217 & 4819 \\
ResiClust &  &  & 11738 & 4663 \\
RiFuClust &  &  &  & 10245 \\
\hline
 	\end{tabular}
\end{center}
\label{tab:compare}
\end{table}

\begin{table}
\caption{Analysis of stomach cancer data: regression coefficients for gene expression TGFBR2.}
 \begin{center}
    \setlength{\tabcolsep}{3.5pt}
    \renewcommand{\arraystretch}{1.0}
	\begin{tabular}{lcccc} \hline
 & Subgroup 1 & Subgroup 2 & Subgroup 3 & Subgroup 4 \\ \hline
 FGF14 & -0.372 &  &  &  \\
TGFB3 & -0.355 &  &  &  \\
DAXX & -0.304 &  &  &  \\
BRAF & -0.255 &  &  &  \\
HRAS &  & 0.236 & 0.155 &  \\
PLA2G4E &  & 0.198 &  &  \\
DUSP1 &  &  &  & -0.102 \\
DUSP2 &  & 0.106 &  &  \\
DUSP16 &  &  &  & 0.102 \\
TGFBR2 &-0.349  & -0.081 & -0.303 & -0.371 \\
KRAS &  & 0.016 &  &  \\
FLNC &  &  & -0.142 &  \\
PAK2 &  &  & 0.122 &  \\
FGF7 &  &  & -0.106 &  \\
TRAF2 &  &  &  & 0.155 \\
PPP3CC &  &  &  & -0.086 \\
\hline
 	\end{tabular}
\end{center}
\label{tab:stomach_coef}
\end{table}

\renewcommand\thefigure{C\arabic{figure}}
\setcounter{figure}{0}

\begin{figure}[htbp]
   \centering
   \includegraphics[scale=0.5]{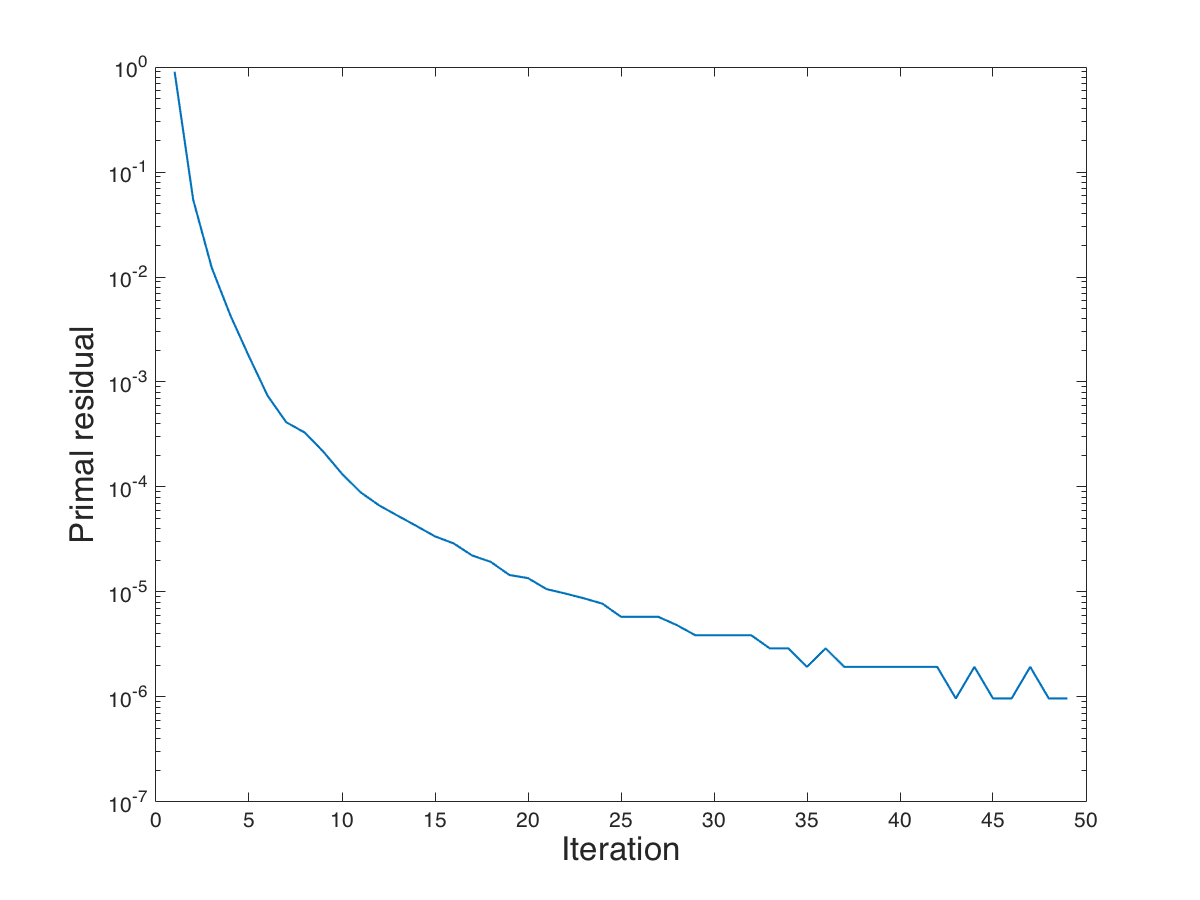}
   \caption{Simulation: primal residual as a function of iteration for one replicate under Setting S1, balanced structure, $(\rho,\sigma^2)=(0.3,1)$}
   \label{fig:time}
\end{figure}

\begin{figure}[htbp]
   \centering
   \includegraphics[scale=0.35]{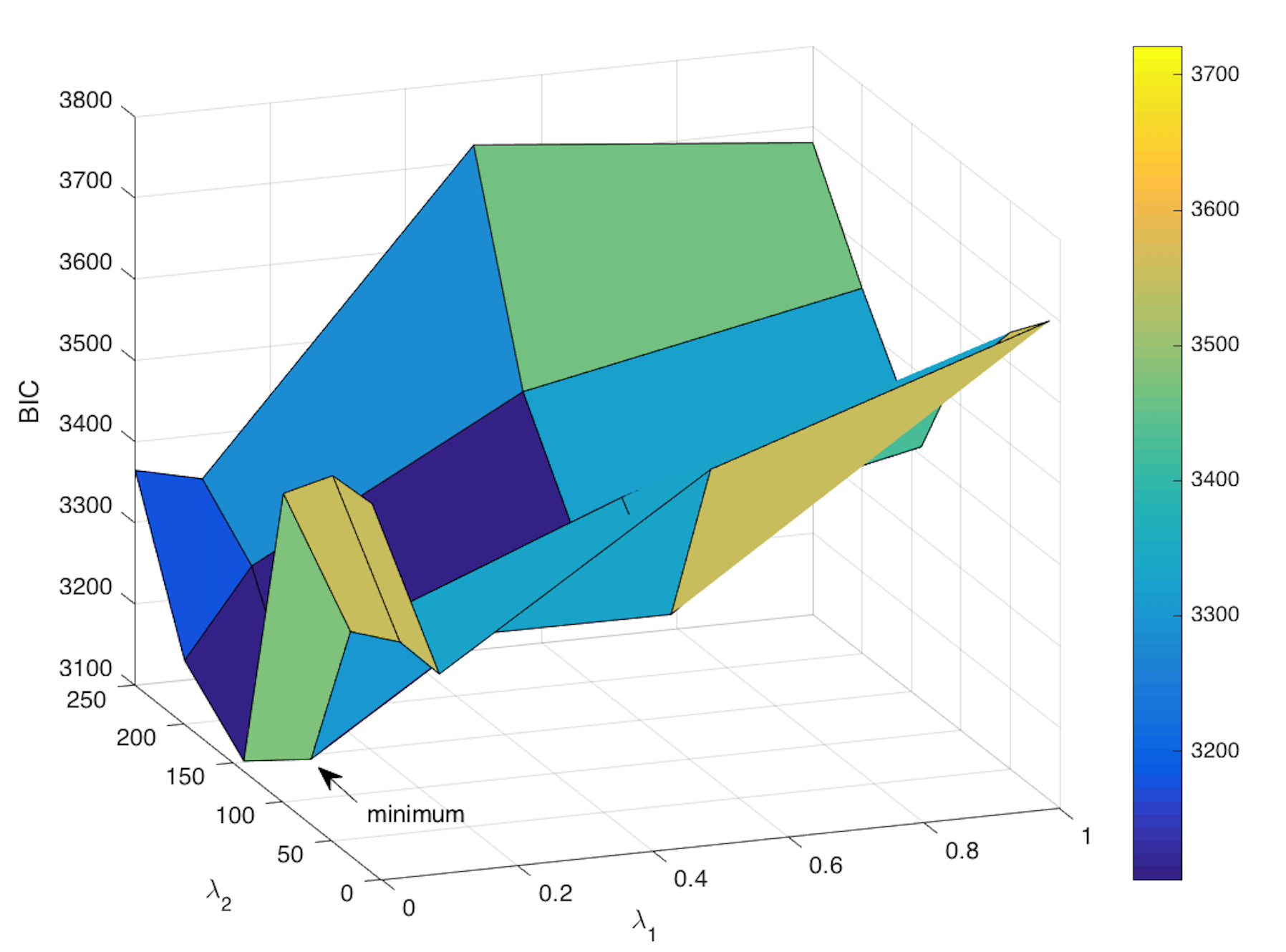}
   \caption{Simulation: BIC as a function of $\lambda_1$ and $\lambda_2$ for one replicate under setting S1, balanced structure, $(\rho,\sigma^2)=(0.3,1)$. }
   \label{fig:BIC}
\end{figure}

\begin{figure}[htbp]
   \centering
   \includegraphics[scale=0.7]{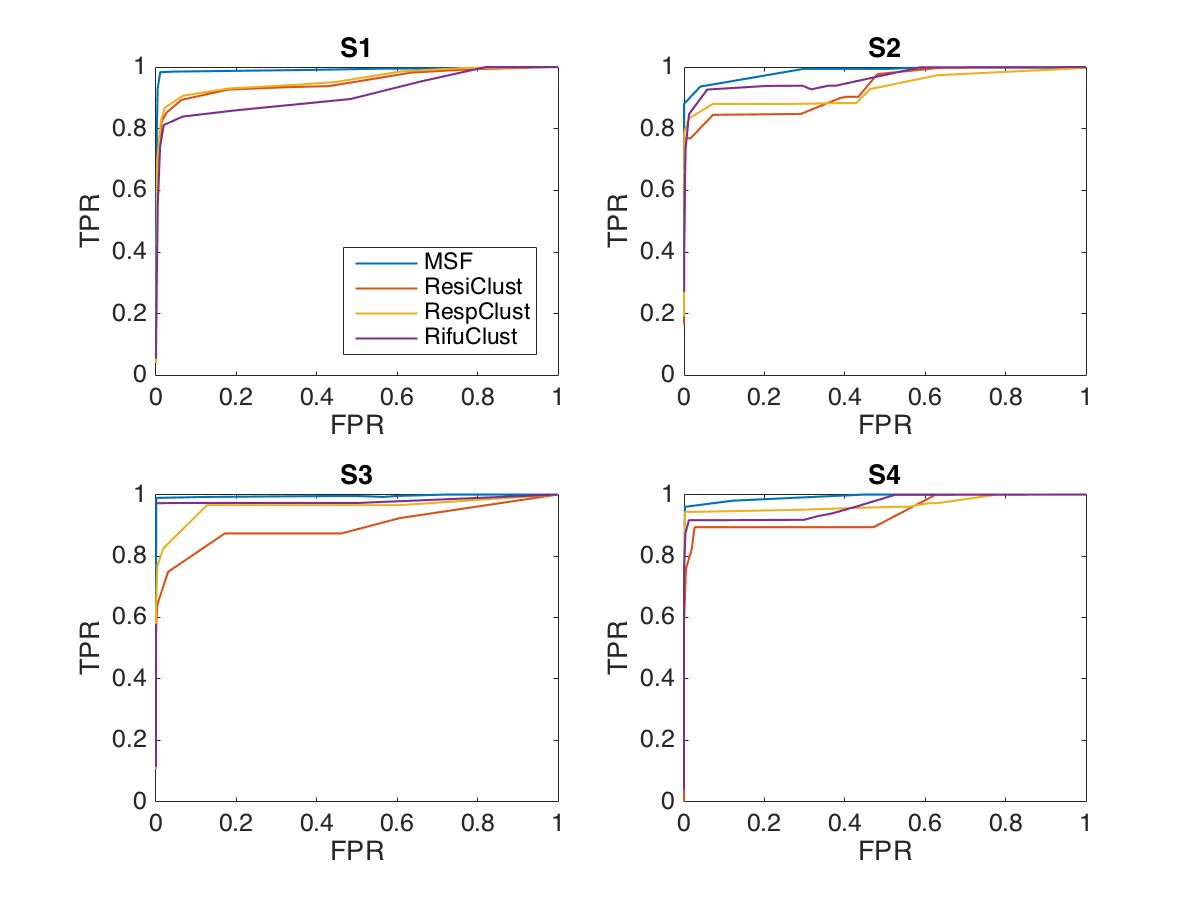}
   \caption{Simulation: ROC curves for variable selection. Under settings S1 and S2 (first row), data are generated with a balanced subgroup structure and $(\rho,\sigma^2)=(0.3,1)$. Under settings S3 and S4 (second row), data are generated with a balanced subgroup structure and $(\rho,\phi)=(0.3,0.3)$.}
   \label{fig:roc}
\end{figure}

\begin{figure}[htbp]
   \centering
   \includegraphics[scale=0.2]{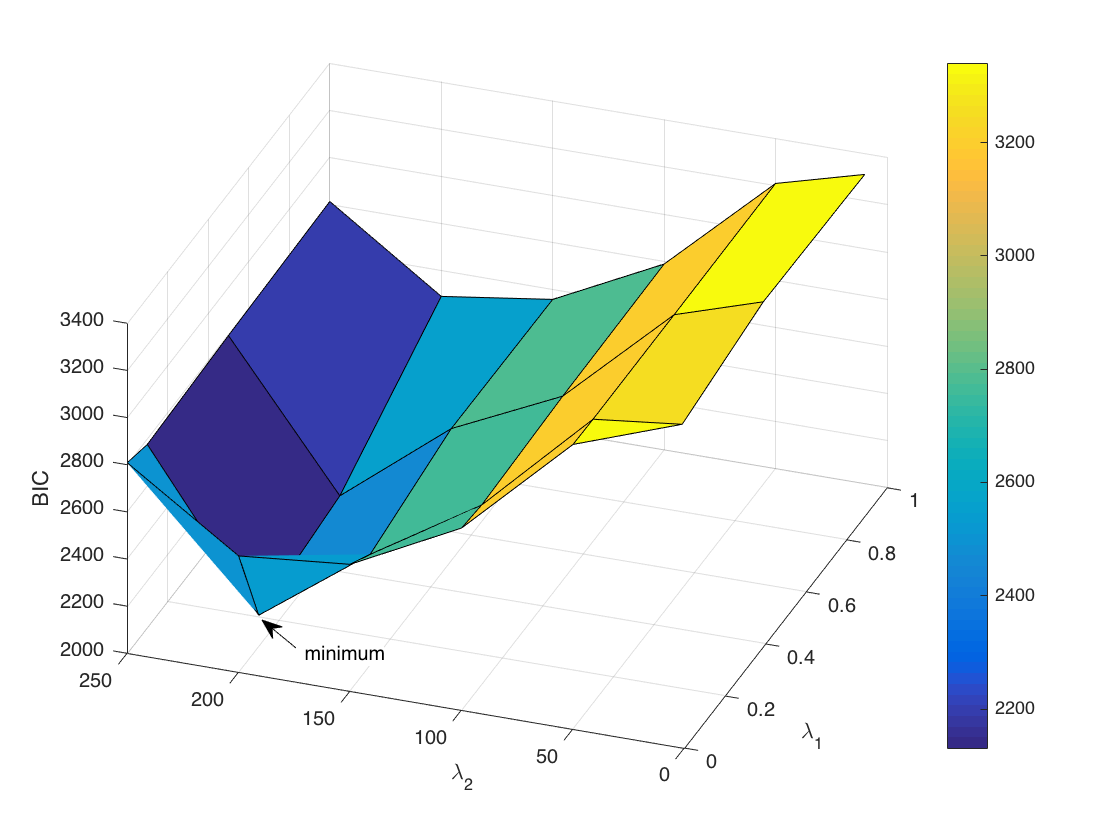}
    \includegraphics[scale=0.2]{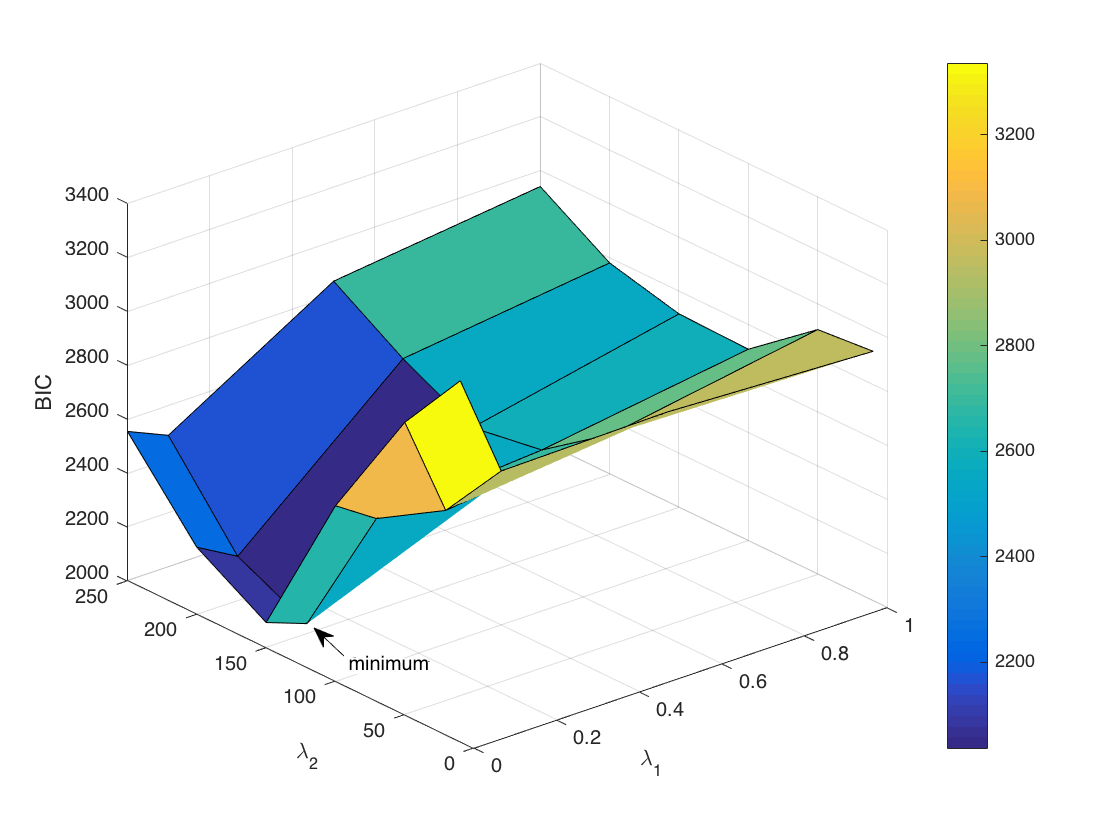}
   \caption{Data analysis: BIC as a function of $\lambda_1$ and $\lambda_2$. Left: SKCM. Right: stomach cancer data.}
   \label{fig:BIC_data}
\end{figure}


\begin{figure}[htbp]
   \centering
   \includegraphics[scale=0.2]{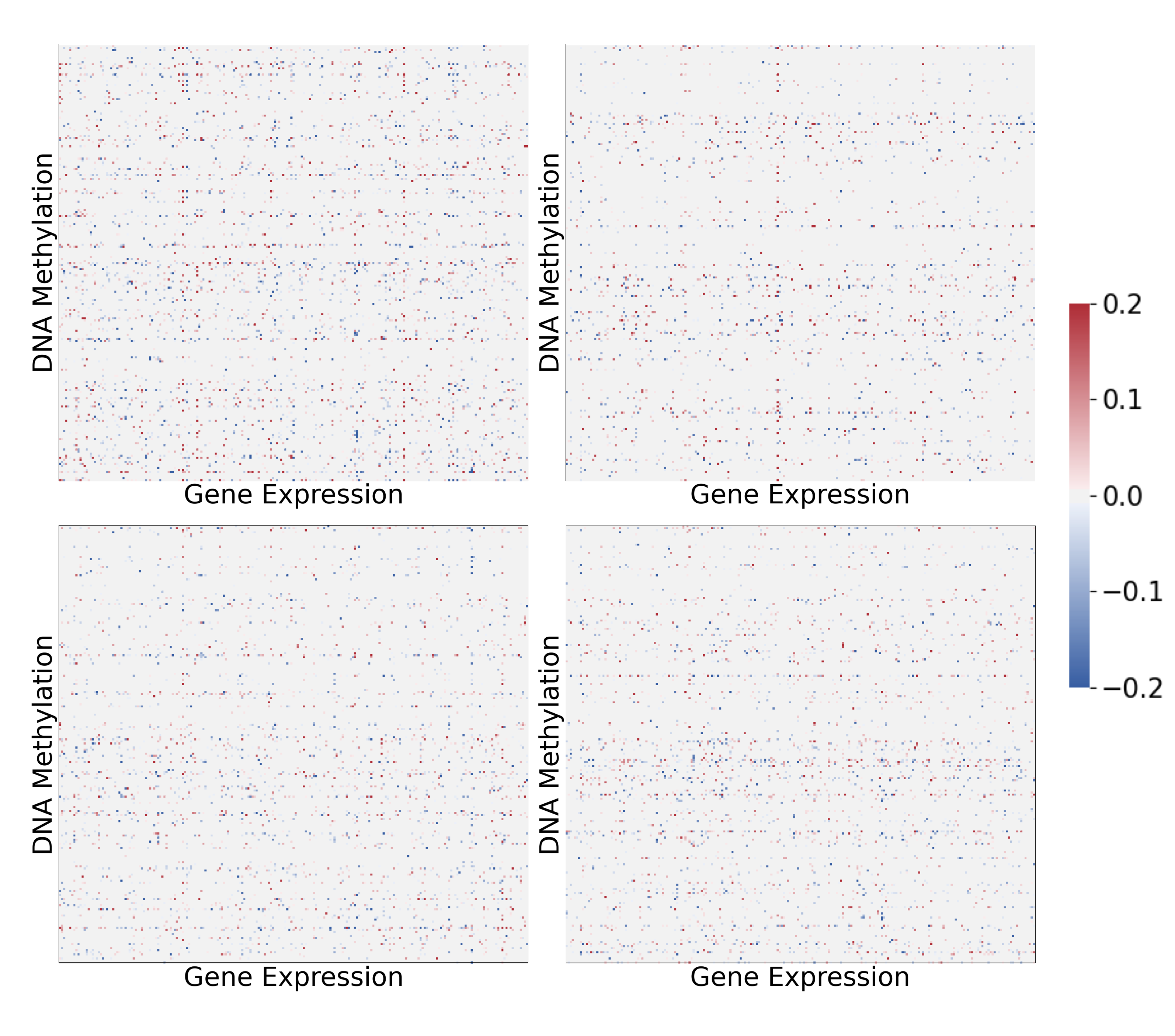}
   \caption{Analysis of stomach cancer data: heatmap of the estimated regulation coefficient matrix. Each panel corresponds to one subgroup. {\color{blue}The four coefficient matrices are significantly different ($p$-values<$0.001$).}}
   \label{fig:stomach.heat}
\end{figure}

\end{document}